\documentclass[twocolumn,showpacs,preprintnumbers,superscriptaddress,amsmath,amssymb,pre]{revtex4-1}

\usepackage{graphicx,easymat,chemarrow}
\usepackage{dcolumn}
\usepackage{bm}
\usepackage{color}
\usepackage{multirow}
\usepackage{listings}

\begin{document}

\preprint{PRE/MF-X-DMA}

\title{Multifractal detrending moving average cross-correlation analysis}

\author{Zhi-Qiang Jiang}
 \affiliation{School of Business, East China University of Science and Technology, Shanghai 200237, China} %
 \affiliation{Research Center for Econophysics, East China University of Science and Technology, Shanghai 200237, China} %

\author{Wei-Xing Zhou}
 \email{wxzhou@ecust.edu.cn}
 \affiliation{School of Business, East China University of Science and Technology, Shanghai 200237, China} %
 \affiliation{Research Center for Econophysics, East China University of Science and Technology, Shanghai 200237, China} %
 \affiliation{School of Science, East China University of Science and Technology, Shanghai 200237, China} %

\date{\today}

\begin{abstract}
There are a number of situations in which several signals are simultaneously recorded in complex systems, which exhibit long-term power-law cross-correlations. The multifractal detrended cross-correlation analysis (MF-DCCA) approaches can be used to quantify such cross-correlations, such as the MF-DCCA based on detrended fluctuation analysis (MF-X-DFA) method. We develop in this work a class of MF-DCCA algorithms based on the detrending moving average analysis, called MF-X-DMA. The performances of the MF-X-DMA algorithms are compared with the MF-X-DFA method by extensive numerical experiments on pairs of time series generated from bivariate fractional Brownian motions, two-component autoregressive fractionally integrated moving average processes and binomial measures, which have theoretical expressions of the multifractal nature. In all cases, the scaling exponents $h_{xy}$ extracted from the MF-X-DMA and MF-X-DFA algorithms are very close to the theoretical values. For bivariate fractional Brownian motions, the scaling exponent of the cross-correlation is independent of the cross-correlation coefficient between two time series and the MF-X-DFA and centered MF-X-DMA algorithms have comparative performance, which outperform the forward and backward MF-X-DMA algorithms. For two-component autoregressive fractionally integrated moving average processes, we also find that the MF-X-DFA and centered MF-X-DMA algorithms have comparative performance while the forward and backward MF-X-DMA algorithms perform slightly worse. For binomial measures, the forward MF-X-DMA algorithm exhibits the best performance, the centered MF-X-DMA algorithms performs worst, and the backward MF-X-DMA algorithm outperforms the MF-X-DFA algorithm when the moment order $q<0$ and underperforms when $q>0$. We apply these algorithms to the return time series of two stock market indexes and to their volatilities. For the returns, the centered MF-X-DMA algorithm gives the best estimates of $h_{xy}(q)$ since its $h_{xy}(2)$ is closest to 0.5 as expected, and the MF-X-DFA algorithm has the second best performance. For the volatilities, the forward and backward MF-X-DMA algorithms give similar results, while the centered MF-X-DMA and the MF-X-DFA algorithms fails to extract rational multifractal nature.
\end{abstract}


\maketitle


\section{Introduction}
\label{sec:Introduction}

Natural and socio-economic systems are usually complex systems from which macroscopic statistical laws emerge. These macroscopic laws are the outcomes of self-organization and interactions among constituents through, which cannot be explained by the sum of the microscopic behaviors of individuals. Statistical laws can be extracted from time series, which are the most usual records of observable quantities in real world. The fractal and multifractal nature of time series have been extensively studies for different systems \cite{Mandelbrot-1983}.

For a nonstationary time series, the detrended fluctuation analysis (DFA) can be adopted to explore its long-range autocorrelations \cite{Peng-Buldyrev-Havlin-Simons-Stanley-Goldberger-1994-PRE,Hu-Ivanov-Chen-Carpena-Stanley-2001-PRE,Chen-Ivanov-Hu-Stanley-2002-PRE,Chen-Hu-Carpena-Bernaola-Galvan-Stanley-Ivanov-2005-PRE,Ma-Bartsch-BernaolaGalvan-Yoneyama-Ivanov-2010-PRE} and multifractal features \cite{CastroESilva-Moreira-1997-PA,Weber-Talkner-2001-JGR,Kantelhardt-Zschiegner-KoscielnyBunde-Havlin-Bunde-Stanley-2002-PA}. Alternatively, the detrending moving average (DMA) method can also be used for fractal analysis \cite{Vandewalle-Ausloos-1998-PRE,Alessio-Carbone-Castelli-Frappietro-2002-EPJB,AlvarezRamirez-Rodriguez-Echeverria-2005-PA,Arianos-Carbone-2007-PA,Carbone-2009-IEEE} or multifractal analysis \cite{Gu-Zhou-2010-PRE}. Numerical experiments on monofractal time series unveil that the performance of the DMA method is comparable to the DFA method with slightly different priorities under different situations \cite{Grech-Mazur-2005-APPB,Xu-Ivanov-Hu-Chen-Carbone-Stanley-2005-PRE,Bashan-Bartsch-Kantelhardt-Havlin-2008-PA}. However, for multifractal time series, the multifractal detrending moving average (MFDMA) performs better than the multifractal detrended fluctuation analysis (MFDFA) \cite{Gu-Zhou-2010-PRE}. In addition, we note that both the DFA and DMA algorithms can be extended from one dimension to higher dimensions for fractal and multifractal analysis \cite{Gu-Zhou-2006-PRE,Carbone-2007-PRE,Turk-Carbone-Chiaia-2010-PRE,Gu-Zhou-2010-PRE}.

A complex system usually contains several observable variables that exhibit long-range dependence or multifractal nature. In turbulent flows, the velocity, temperature and concentration fields are embedded in the same space as joint multifractal measures
\cite{Antonia-VanAtta-1975-JFM,Meneveau-Sreenivasan-Kailasnath-Fan-1990-PRA,Schmitt-Schertzer-Lovejoy-Brunet-1996-EPL,Xu-Antonia-Rajagopalan-2000-EPL,Xu-Antonia-Rajagopalan-2007-EPL}, in which the scaling behavior of the joint moments of two joint multifractal measures $\mu_1$ and $\mu_2$ are investigated
\begin{equation}
 J(s) = \langle[\mu_1(s)]^p [\mu_2(s)]^q\rangle,
 \label{Eq:MFDCCA:Jr}
\end{equation}
where $s$ is the box size. This framework has also been applied to study the joint multifractal nature between topographic indices and crop yield in agronomy \cite{Kravchenko-Bullock-Boast-2000-AJ,Zeleke-Si-2004-AJ}, trading volume and volatility in stock markets \cite{Lin-2008-PA}, nitrogen dioxide and ground-level ozone \cite{JimenezHornero-JimenezHornero-deRave-PavonDominguez-2010-EMA},
heart rate variability and brain activity of healthy humans \cite{Lin-Sharif-2010-Chaos}, and wind patterns and land surface air temperature \cite{JimenezHornero-PavonDominguez-deRave-ArizaVillaverde-2010-AR}.

For two stationary time series $\{x(i)\}$ and $\{y(i)\}$ of the same length, the time-lagged cross-correlation or covariance provides another example \cite{Lavancier-Philippe-Surgailis-2009-SPL,Coeurjolly-Amblard-Achard-2010-EUSIPCO,Amblard-Coeurjolly-Lavancier-Philippe-2011-BSMF,Podobnik-Wang-Horvatic-Grosse-Stanley-2010-EPL},
\begin{equation}
 C(s) = \langle{x(t)y(t+s)}\rangle.
 \label{Eq:MFDCCA:Cs:1}
\end{equation}
For two nonstationary time series $\{x(i)\}$ and $\{y(i)\}$ of the same length, one can study the following cross-correlation function between two detrended series \cite{Arianos-Carbone-2009-JSM}:
\begin{equation}
 C_{xy}(s) = \left\langle{\left[X(t)-\widetilde{X}(t)\right]\left[Y(t+s)-\widetilde{Y}(t+s)\right]}\right\rangle,
 \label{Eq:MFDCCA:Cs:2}
\end{equation}
where $\widetilde{X}(t)$ and $\widetilde{Y}(t)$ are certain trend functions of $X(t)$ and $Y(t)$, respectively. The detrended cross-correlation analysis (DCCA) was introduced to investigate the long-range power-law cross-correlations between two nonstationary time series \cite{Jun-Oh-Kim-2006-PRE,Podobnik-Stanley-2008-PRL}:
\begin{equation}
 F_{xy}(s) = \left\langle{\left[X(t)-\widetilde{X}(t)\right]\left[Y(t)-\widetilde{Y}(t)\right]}\right\rangle,
 \label{Eq:MFDCCA:Fs}
\end{equation}
where $\widetilde{X}(t)$ and $\widetilde{Y}(t)$ are certain trend functions of $X(t)$ and $Y(t)$ specific to moving windows of size $s$, respectively. The DFA method is a special case of this DCCA method when $X(t)=Y(t)$. The DCCA method studies the temporal (not the cross-sectional) properties of two nonstationary time series, which is similar to the instant cross-correlations \cite{Qiu-Zheng-Chen-2010-NJP,Qiu-Chen-Zhong-Lei-2011-PA}. The significance of the cross-correlation can be assessed by statistical tests \cite{Podobnik-Grosse-Horvatic-Ilic-Ivanov-Stanley-2009-EPJB,Zebende-2011-PA}. The DCCA method has been applied to study
volume change and price change of the Standard and Poor's (S\&P) 500 Index \cite{Podobnik-Horvatic-Petersen-Stanley-2009-PNAS}, volatilities of the Brazilian agrarian commodities and stocks \cite{SiqueirJr-Stosic-Bejan-Stosic-2010-PA}, traffic flows \cite{Xu-Shang-Kamae-2010-ND}, and self-affine time series of taxi accidents \cite{Zebende-daSilva-Filho-2011-PA}.

More generally, the multifractal detrneded cross-correlation analysis was introduced to investigate the multifractal nature in the long-range power-law cross-correlations between two nonstationary time series \cite{Zhou-2008-PRE}, which recovers the MFDFA method when $X(t)=Y(t)$. We call this method the MF-X-DFA for the reason that will be clear in Sec.~\ref{S1:Algo:MF-X-DMA}. Note that the MF-X-DFA method is relevant to the multifractal height cross-correlation analysis with differences \cite{Kristoufek-2010}. The MF-X-DFA method has been applied to temporal and spatial seismic data \cite{Shadkhoo-Jafari-2009-EPJB}, sunspot numbers and river flow fluctuations \cite{Hajian-Movahed-2010-PA}, stock index returns \cite{Wang-Wei-Wu-2010-PA,Sun-Sheng-2010-BIFE}, price-volume relationships in agricultural commodity futures markets \cite{He-Chen-2011a-PA}, and spot and futures markets of WTI crude oil \cite{Wang-Wei-Wu-2011-PA}.

In this work, we introduce a variant of the MF-X-DFA algorithm, termed multifractal detrending moving average cross-correlation analysis (MF-X-DMA), which combines the ideas of MFDMA and DCCA. The main difference between MF-X-DFA and MF-X-DMA is that the latter adopts local moving average as the trend function. Since the MFDMA algorithm outperforms the MFDFA algorithm for multifractal time series, we expect that the MF-X-DMA algorithm will show advantages over the MF-X-DFA algorithm. Our numerical experiments and real-work data analysis confirm this conjecture.

The paper is organized as follows. The paper is organized as follows. Section \ref{S1:Algo:MF-X-DMA} describes a unified framework of the MF-X-DFA and MF-X-DMA algorithms. Section \ref{S1:NumSim} performs extensive numerical experiments using fractal and multifractal time series with known analytical expressions (bivariate fractional Brownian motions, two-component ARFIMA processes and binomial measures) to investigate the performance of the algorithms. In Sec.~\ref{S1:Application}, we apply the algorithms to daily stock index returns and volatilities. We discuss and summarize our findings in Sec.~\ref{S1:conclusion}.

\section{MF-X-DMA and MF-X-DFA}
\label{S1:Algo:MF-X-DMA}

Consider two stationary time series $\{x(i)\}$ and $\{y(i)\}$ of the same length $M$, where $i=1, 2, \cdots, M$. Without loss of generality, we can assume that these two time series have zero means. Each time series is covered with $M_s=[M/s]$ non-overlapping boxes of size $s$. The profile within the $v$th box $[l_v+1,l_v+s]$, where $l_v=(v-1)s$, are determined to be
\begin{equation}
 X_v(k) = \sum_{j=1}^{k} x(l_v+j) {\rm{~and~}} Y_v(k) = \sum_{j=1}^{k} y(l_v+j),
\end{equation}
where $k=1,\cdots,s$. Assume that the local trending functions of $\{X_v(k)\}$ and $\{Y_v(k)\}$ are $\{\widetilde{X}_v(k)\}$ and $\{\widetilde{Y}_v(k)\}$, respectively. The cross-correlation for each box is calculated as follows
\begin{equation}
 F_{v}(s) = \frac{1}{s}\sum_{k=1}^s \left[X_v(k)-\widetilde{X}_v(k)\right]\left[Y_v(k)-\widetilde{Y}_v(k)\right]
\end{equation}
The $q$th order cross-correlation is calculated as follows
\begin{equation}
 F_{xy}(q,s) = \left[\frac{1}{m}\sum_{v=1}^m |F_{v}(s)|^{q/2}\right]^{1/q}
\end{equation}
when $q\neq0$ and
\begin{equation}
 F_{xy}(0,s) = \exp\left[\frac{1}{2m}\sum_{v=1}^m \ln |F_{v}(s)|\right]~.
\end{equation}
We then expect the following scaling relation
\begin{equation}
 F_{xy}(q,s) \sim s^{h_{xy}(q)}~.
 \label{Eq:Fxy:s}
\end{equation}

According to the standard multifractal formalism, the multifractal scaling exponent $\tau(q)$ can be used to characterize the multifractal nature, which reads
\begin{equation}
\tau_{xy}(q)=qh_{xy}(q)-D_f,
\label{Eq:MFDCCA:tau}
\end{equation}
where $D_f$ is the fractal dimension of the geometric support of the multifractal measure \cite{Kantelhardt-Zschiegner-KoscielnyBunde-Havlin-Bunde-Stanley-2002-PA}. For time series analysis, we have $D_f=1$. If the scaling exponent function $\tau(q)$ is a nonlinear function of $q$, the signal has multifractal nature. It is easy to obtain the singularity strength function $\alpha(q)$ and the multifractal spectrum $f(\alpha)$ via the Legendre transform \cite{Halsey-Jensen-Kadanoff-Procaccia-Shraiman-1986-PRA}
\begin{equation}
    \left\{
    \begin{array}{ll}
        \alpha_{xy}(q)={\rm{d}}\tau_{xy}(q)/{\rm{d}}q\\
        f_{xy}(q)=q{\alpha_{xy}}-{\tau_{xy}}(q)
    \end{array}
    \right..
\label{Eq:MFDCCA:f:alpha}
\end{equation}

There are many different methods for the determination of $\widetilde{X}_v$ and $\widetilde{Y}_v$. The local detrending functions could be polynomials \cite{Peng-Buldyrev-Havlin-Simons-Stanley-Goldberger-1994-PRE,Hu-Ivanov-Chen-Carpena-Stanley-2001-PRE}, which recovers the MF-DXA method \cite{Zhou-2008-PRE}. The local detrending function could also be the moving averages \cite{Vandewalle-Ausloos-1998-PRE,Alessio-Carbone-Castelli-Frappietro-2002-EPJB}, in which the algorithm is called MF-X-DMA. To be more clear, we rename the MF-DXA algorithm as MF-X-DFA, and all multifractal analysis algorithms for cross-correlations based on local detrending are termed multifractal detrending/detrended cross-correlation analysis (or MF-DCCA as abbreviation). MF-X-DFA is an MF-DCCA method based on DFA and MF-X-DMA is an MF-DCCA method based on DMA. When $X=Y$, MF-X-DFA in Ref.~\cite{Zhou-2008-PRE} reduces to MF-DFA in Ref.~\cite{Kantelhardt-Zschiegner-KoscielnyBunde-Havlin-Bunde-Stanley-2002-PA}, and MF-X-DMA reduces to MF-DMA in Ref.~\cite{Gu-Zhou-2010-PRE}. We note that the extension of the MF-DCCA algorithms to high dimensions is straightforward \cite{Zhou-2008-PRE}.

The moving average function $\widetilde{Z}(t)$ of $Z\in\{X,Y\}$ in a moving window \cite{Arianos-Carbone-2007-PA} can be calculated as follows,
\begin{equation}
  \widetilde{Z}(t)=\frac{1}{n}\sum_{k=-\lfloor(n-1)\theta\rfloor}^{\lceil(n-1)(1-\theta)\rceil}Z(t-k),
  \label{Eq:MA}
\end{equation}
where $n$ is the window size, $\lfloor{g}\rfloor$ is the largest integer smaller than $g$, $\lceil{g}\rceil$ is the smallest integer larger than $g$, and $\theta$ is the position parameter with the value varying in the range $[0,1]$. Hence, the moving average function considers $\lceil(n-1)(1-\theta)\rceil$ data points in the past and $\lfloor(n-1)\theta\rfloor$ points in the future. We consider three special cases in this paper. The first case $\theta=0$ refers to the backward moving average \cite{Xu-Ivanov-Hu-Chen-Carbone-Stanley-2005-PRE}, in which the moving average function $\widetilde{Z}(t)$ is calculated over all the past $n-1$ data points of the signal. The second case $\theta=0.5$ corresponds to the centered moving average \cite{Xu-Ivanov-Hu-Chen-Carbone-Stanley-2005-PRE}, where $\widetilde{Z}(t)$ contains half past and half future information in each window. The third case $\theta=1$ is called the forward moving average, where $\widetilde{Z}(t)$ considers the trend of $n-1$ data points in the future.

\begin{figure*}[htb]
\centering
\includegraphics[width=5.5cm]{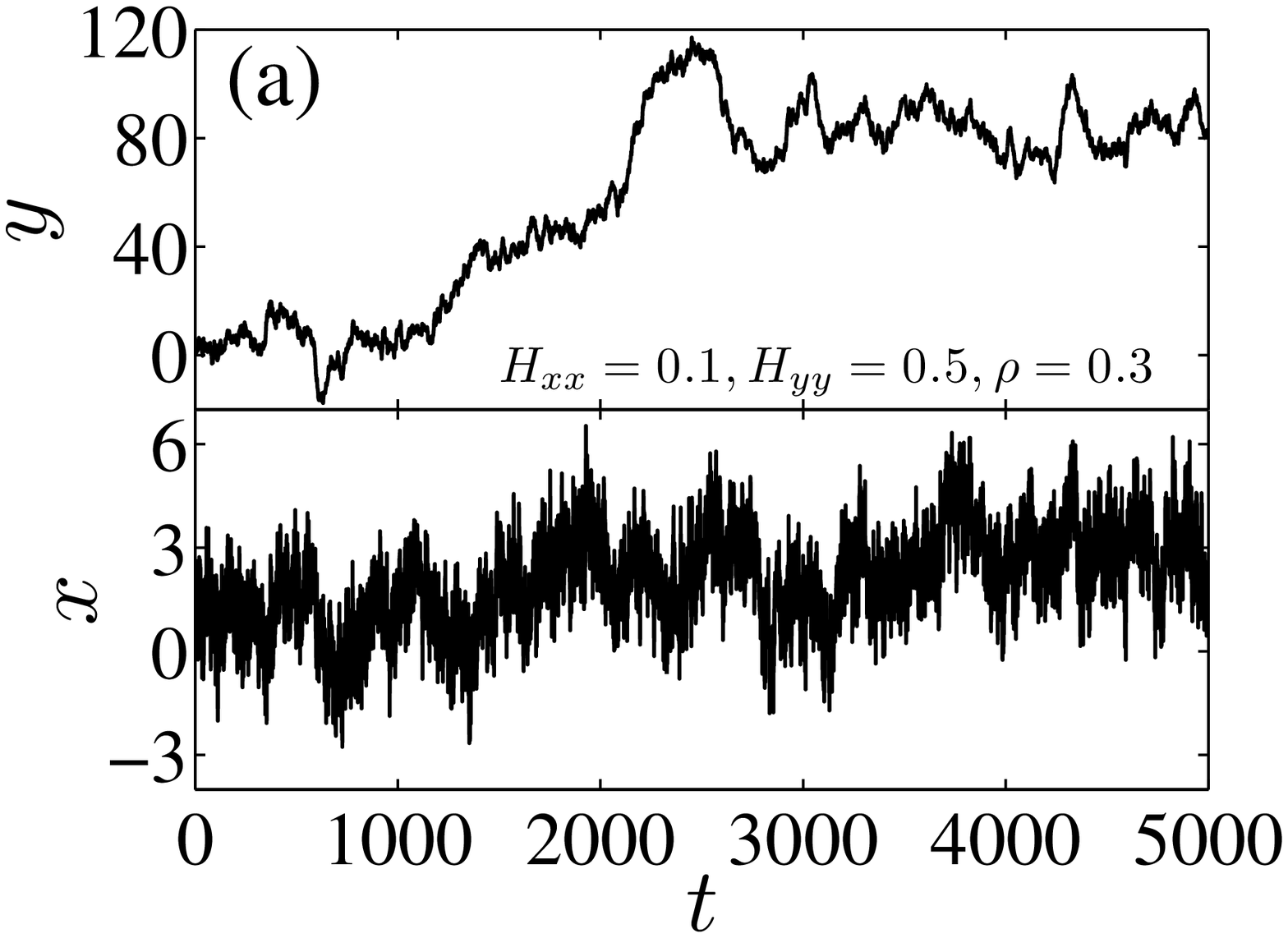}
\includegraphics[width=5.5cm]{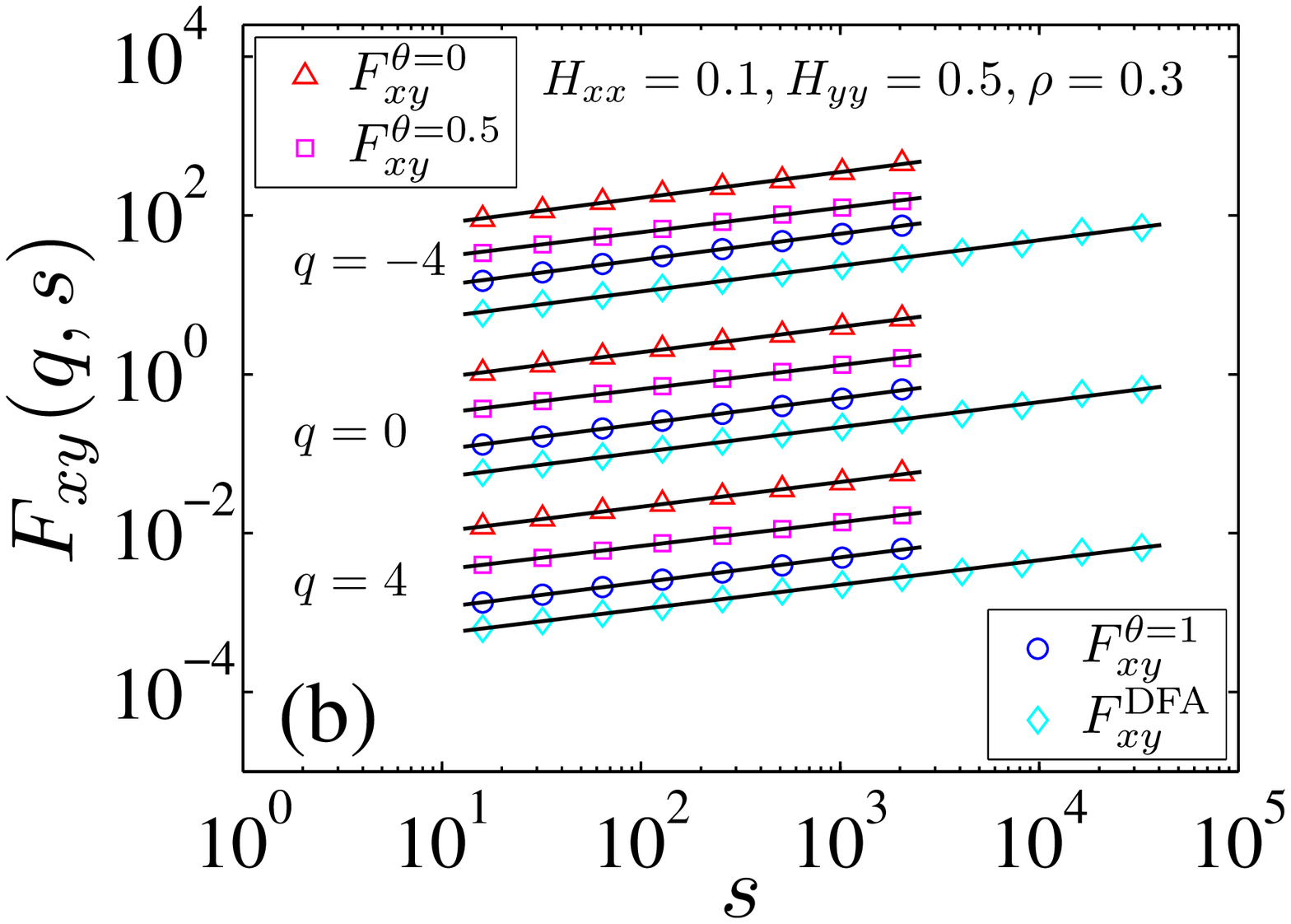}
\includegraphics[width=5.5cm]{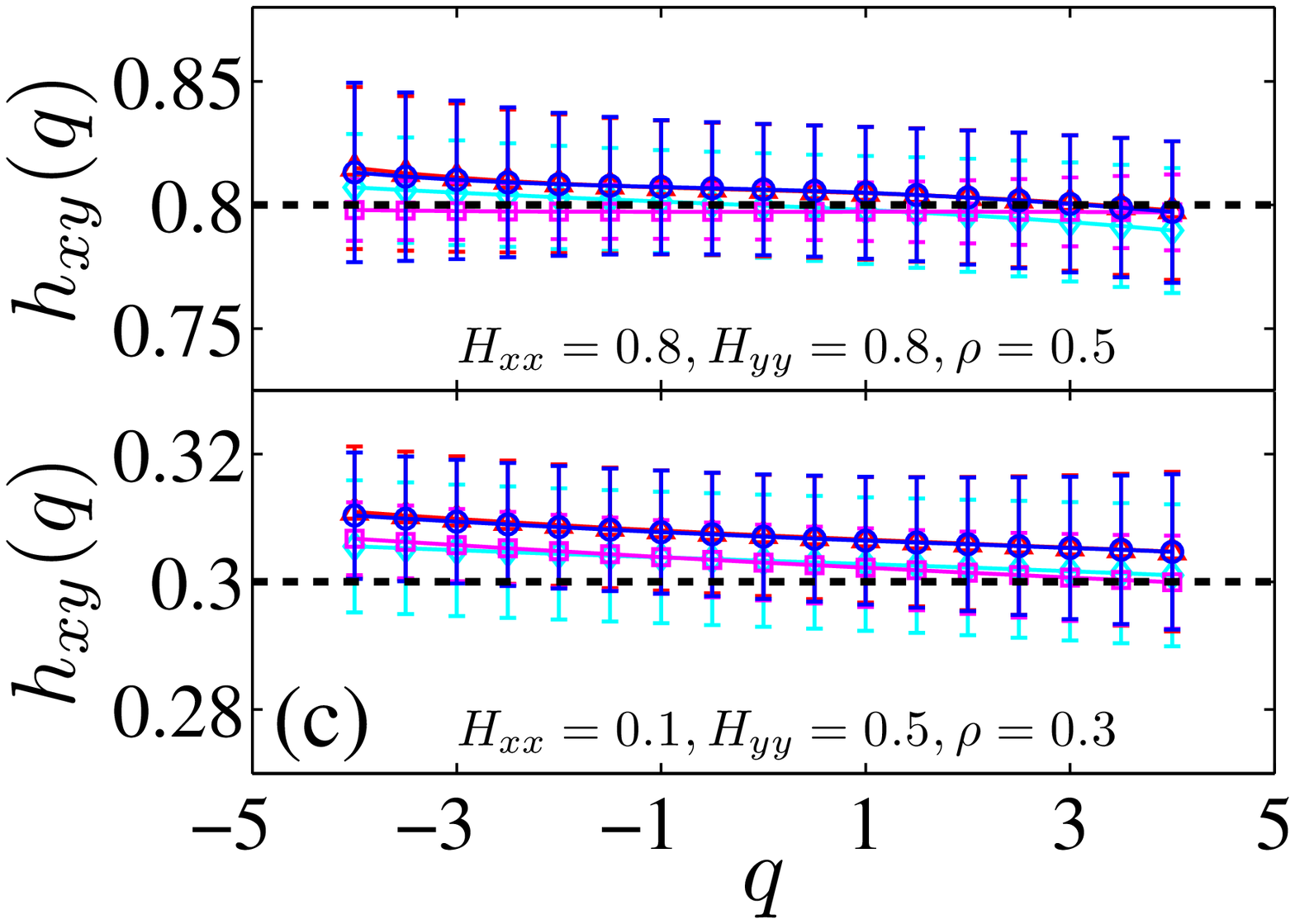}
\includegraphics[width=5.5cm]{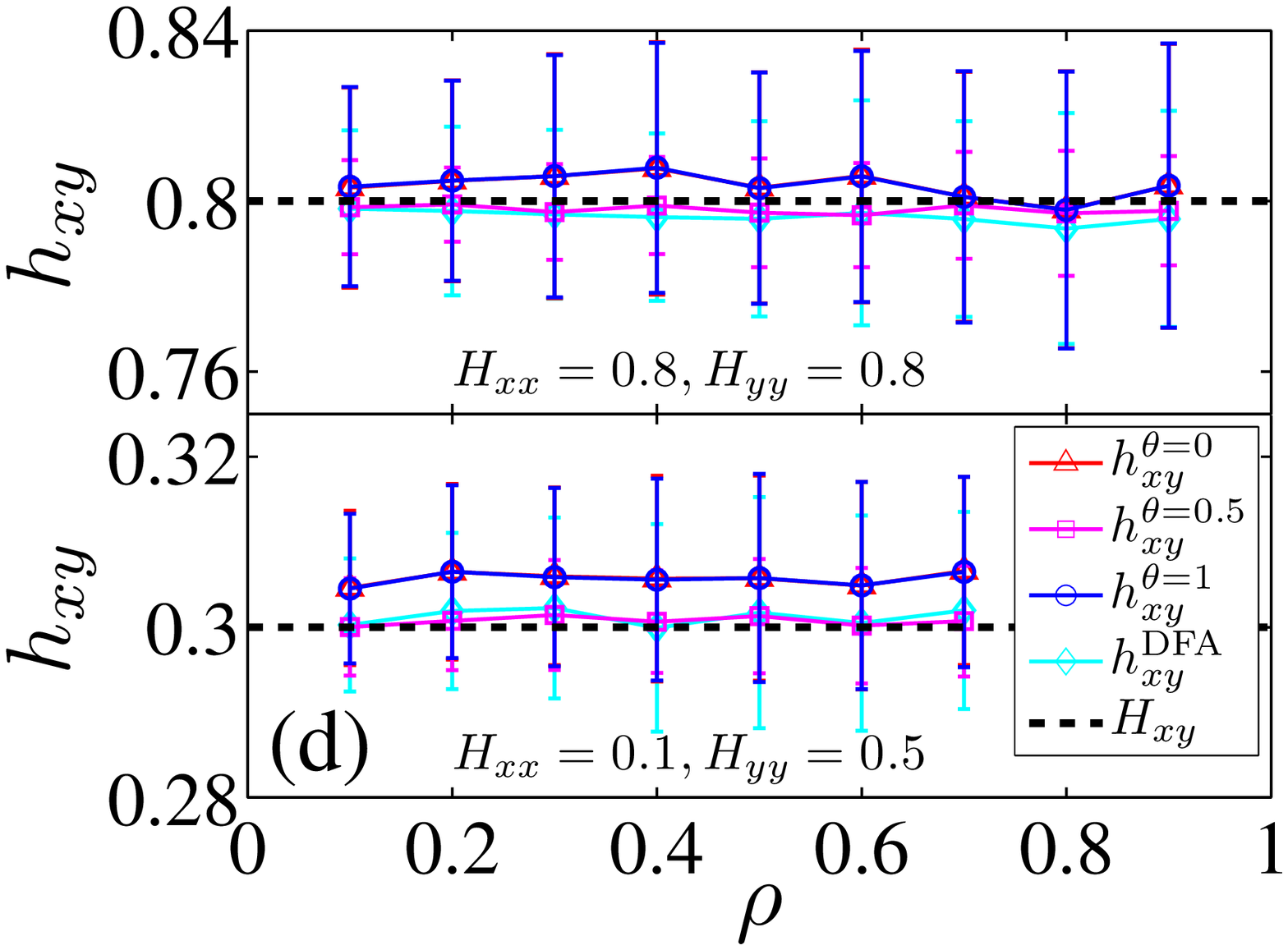}
\includegraphics[width=5.5cm]{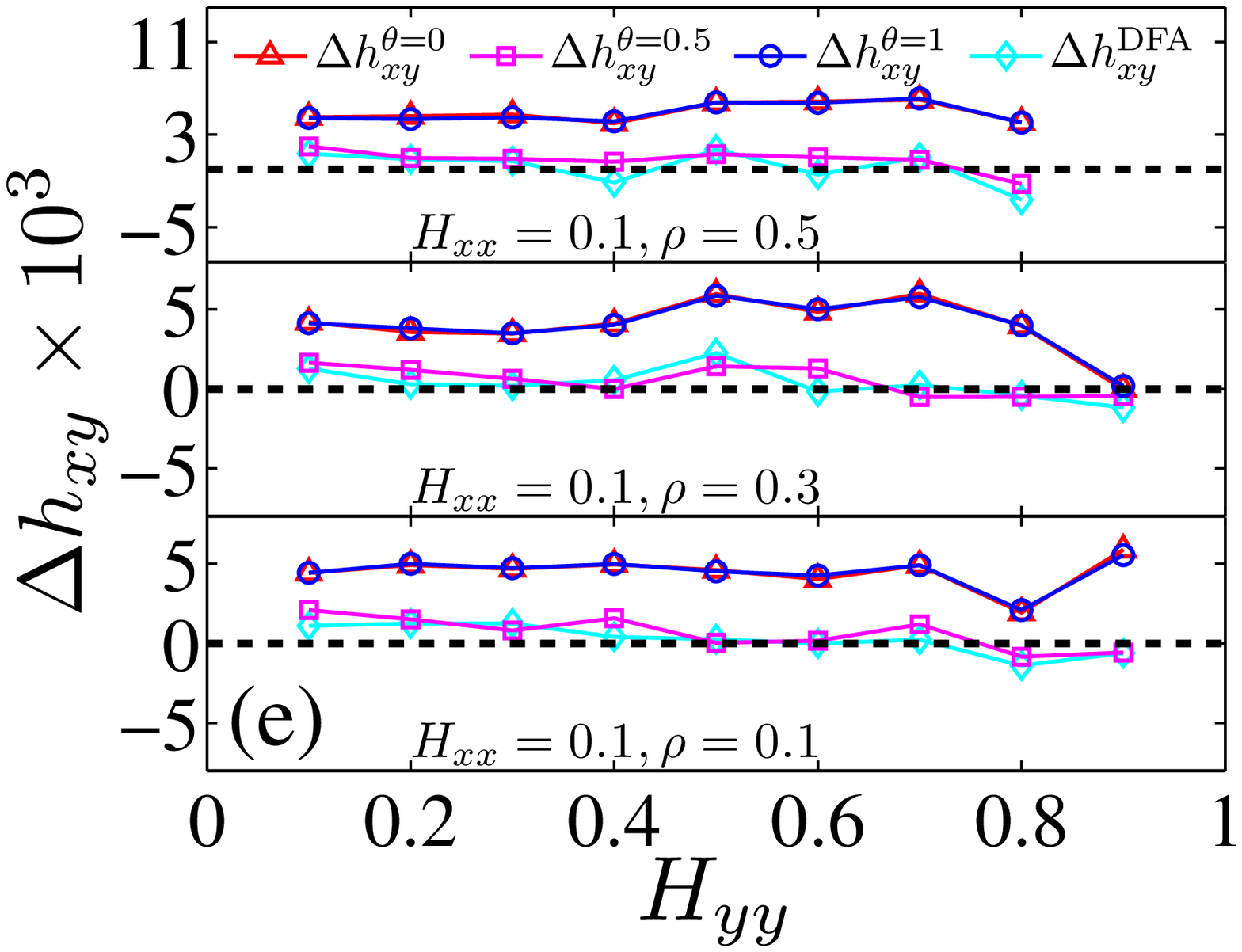}
\includegraphics[width=5.5cm]{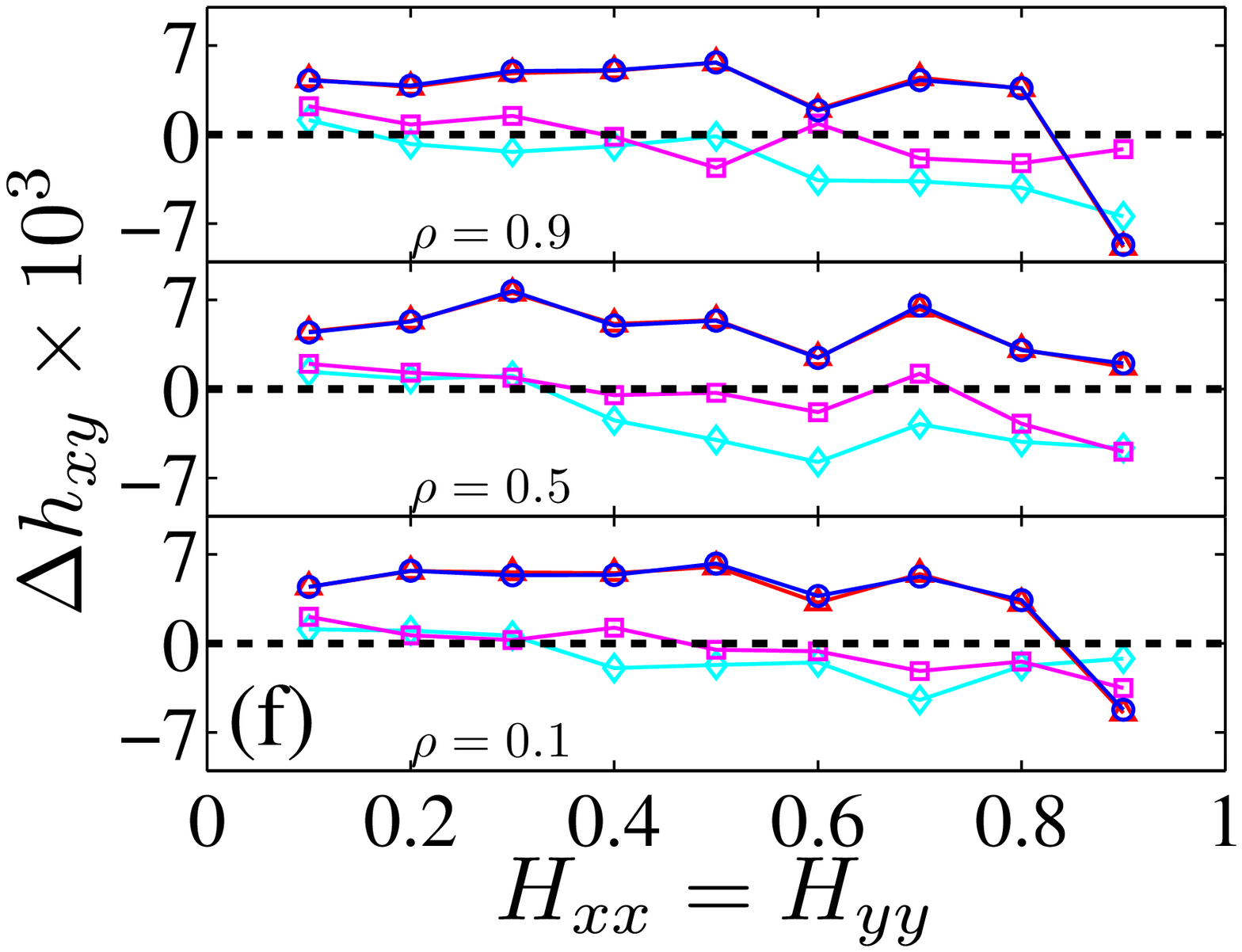}
 \caption{\label{Fig:MFDCCA:bFBM} (Color online) Multifractal detrended cross-correlation analysis of bivariate fractional Brownian motions. Comparisons are performed among three MF-X-DMA algorithms with $\theta=0$, 0.5 and 1 and the MF-X-DFA algorithm. The results in plots (c-f) are averaged over 100 repeated simulations.
 (a) A typical example of bFBM with $H_{xx}=0.1$, $H_{yy}=0.5$ and $\rho=0.3$.
 (b) Power-law dependence of the fluctuation functions $F_{xy}(q,s)$ of the bFBM shown in plot (a) with respect to the scale $s$ for $q=-4$, $q=0$, and $q=4$. The straight lines are the best power-law fits to the data. The results have been translated vertically for better visibility.
 (c) Scaling exponents $h_{xy}(q)$ with the theoretical values as a dashed line for $H_{xx}=H_{yy}=0.8$ and $\rho=0.5$ (up panel) and $H_{xx}=0.1$, $H_{yy}=0.5$ and $\rho=0.3$ (low panel).
 (d) Independence of the scaling exponents $h_{xy}$ with respect to the cross-correlation coefficient $\rho$ for $H_{xx}=H_{yy}=0.8$ (up panel) and $H_{xx}=0.1$ and $H_{yy}=0.5$ (low panel).
 (e) Differences $\Delta{h}_{xy}(q)$ between the estimated scaling exponents $h_{xy}$ and the theoretical exponents $H_{xy}$ for bFBMs, where $H_{xx}=0.1$ is fixed, $H_{yy}$ varies from 0.1 to 0.9, and $\rho$ takes different values.
 (e) Differences $\Delta{h}_{xy}(q)$ between $h_{xy}$ and $H_{xy}$ for bFBMs with $H_{xx}=H_{yy}$ varying from 0.1 to 0.9 and different $\rho$ values.}
\end{figure*}

\section{Numerical experiments}
\label{S1:NumSim}

In order to investigate the validity and performance of the proposed MF-X-DCCA algorithms, we perform extensive numerical experiments using bivariate fractional Brownian motions (bFBMs) \cite{Lavancier-Philippe-Surgailis-2009-SPL,Coeurjolly-Amblard-Achard-2010-EUSIPCO,Amblard-Coeurjolly-Lavancier-Philippe-2011-BSMF}, two-component autoregressive fractionally integrated moving average (ARFIMA) processes \cite{Podobnik-Horvatic-Ng-Stanley-Ivanov-2008-PA,Podobnik-Stanley-2008-PRL}, and binomial measures generated from the multiplicative $p$-model \cite{Meneveau-Sreenivasan-1987-PRL}. By definition, there is no multifractality in bFBMs and two-component ARFIMA processes. Therefore, the $h_{xy}(q)$ function is independent of $q$ and the $\tau_{xy}(q)$ function is linear. In contrast, binomial measures is expected to possess multifractal nature. These three classes of time series are adopted to test the performance of the algorithms since the theoretical expressions of $H_{xx}(q)$ are known for individual time series and we know the theoretical expressions of $H_{xy}(q)$ for the first two classes and the numerical expression for the third class. Note that we have used $H$ for theoretical values and $h$ for estimated values below.

\subsection{Bivariate fractional Brownian motions}

A bivariate fractional Brownian motion $[x(t),y(t)]$ with parameters $\{H_{xx},H_{yy}\}\in(0,1)^2$ is a self-similar Gaussian process with stationary increments, where $x(t)$ and $y(t)$ are two univariate fractional Brownian motions with Hurst indices $H_{xx}$ and $H_{yy}$ and are the two components of the bFBM \cite{Lavancier-Philippe-Surgailis-2009-SPL,Coeurjolly-Amblard-Achard-2010-EUSIPCO,Amblard-Coeurjolly-Lavancier-Philippe-2011-BSMF}. The basic properties of multivariate fractional Brownian motions have been extensively studied \cite{Lavancier-Philippe-Surgailis-2009-SPL,Coeurjolly-Amblard-Achard-2010-EUSIPCO,Amblard-Coeurjolly-Lavancier-Philippe-2011-BSMF}. Especially, it has been proven that the Hurst index $H_{xy}$ of the cross-correlation between the two components is \cite{Lavancier-Philippe-Surgailis-2009-SPL,Coeurjolly-Amblard-Achard-2010-EUSIPCO,Amblard-Coeurjolly-Lavancier-Philippe-2011-BSMF,Arianos-Carbone-2009-JSM}:
\begin{equation}
 H_{xy}=(H_{xx}+H_{yy})/2.
 \label{Eq:Hxy:Hxx:Hyy}
\end{equation}
This property allows us to investigate the performances of the proposed algorithms on a solid foundation.

An efficient simulation technique for univariate FBMs relies upon the embedding of the covariance matrix into a circulant matrix, whose square root can be easily obtained by the discrete Fourier transform \cite{Wood-Chan-1994-JCGS}. This method is an exact simulation algorithm provided that the circulant matrix is definite positive. This algorithm can be generalized to simulate bivariate FBMs, which embeds the circulant of a block Toeplitz covariance matrix and use the fast Fourier transform to diagonalize the block circulant matrix \cite{Chan-Wood-1999-SC}. A detailed description of the simulation procedure can be found in Ref.~\cite{Coeurjolly-Amblard-Achard-2010-EUSIPCO,Amblard-Coeurjolly-Lavancier-Philippe-2011-BSMF}.

In the simulation algorithm, the two Hurst indexes $H_{xx}$ and $H_{yy}$ of the two univariate FBMs and their cross-correlation coefficient $\rho$ are input arguments. We have generated a huge number of bFBMs, where $H_{xx}$, $H_{yy}$ and $\rho$ all vary from 0.1 to 0.9 with a spacing of 0.1. For a given triple of $(H_{xx},H_{yy},\rho)$, 100 repeated simulations are conducted and 100 bFBMs with length of $2^{16}$are generated. In most cases, the definite positivity condition is not fulfilled. We then perform MF-X-DMA and MF-DFA on each bFBM to obtain the scaling exponent $h_{xy}$. The average over 100 repeated simulations is calculated. We have observed for each bFBM and each algorithm that
\begin{equation}
 h_{xy}=(h_{xx}+h_{yy})/2.
 \label{Eq:hxy:hxx:hyy}
\end{equation}
Our main findings are the following: (1) The exponent $h_{xy}$ is independent of the cross-correlation coefficient $\rho$; (2) The $h_{xy}(q)$ functions are independent of $q$, indicating that the bFBMs are monofractals; (3) All the four algorithms give nice estimates $h_{xy}$ of the scaling exponents, which are very close to the corresponding theoretical $H_{xy}$ values; and (4) The centered MF-X-DMA algorithm ($\theta=0.5$) and the MF-X-DFA algorithms have comparative performance and perform better than the backward and forward MF-X-DMA algorithms ($\theta=0$ and $\theta=1$). Since there are too many results to present in a concise way, we present a part of the results to manifest these findings.

A typical example of the bFBM with $H_{xx}=0.1$, $H_{yy}=0.5$ and $\rho=0.3$ is illustrated in Fig.~\ref{Fig:MFDCCA:bFBM}(a) and the corresponding power-law dependence of the fluctuation functions $F_{xy}(q,s)$ with respect to the scale $s$ for the four algorithms is shown in Fig.~\ref{Fig:MFDCCA:bFBM}(b). For MF-X-DMA algorithms, $s$ should not be too large due to the finite-size effect. The scaling ranges span over two orders of magnitude for the MF-X-DMA algorithms and three orders of magnitude for the MF-X-DFA algorithm. In the determination of the scaling exponents $h_{xy}$, we have used the same scaling ranges as in Fig.~\ref{Fig:MFDCCA:bFBM}(b) for all the bFBMs and nice power-law relationships are observed.

The two panels of Fig.~\ref{Fig:MFDCCA:bFBM}(c) show the $h_{xy}(q)$ functions for $H_{xx}=H_{yy}=0.8$ and $\rho=0.5$ and for $H_{xx}=0.1$, $H_{yy}=0.5$ and $\rho=0.3$, respectively. Although there is a decreasing trend in each function, the theoretical functions $H_{xy}(q)=0.8$ and $H_{xy}(q)=0.3$ are well within the error bars, indicating that the $h_{xy}(q)$ functions are independent of the order $q$. Hence, the four algorithms are able to correctly capture the monofractal nature of the bFBMs. We focus on $q=2$ below.

The two panels of Fig.~\ref{Fig:MFDCCA:bFBM}(d) show the dependence of the scaling exponents $h_{xy}$ with respect to the cross-correlation coefficient $\rho$ for $H_{xx}=H_{yy}=0.8$ and for $H_{xx}=0.1$ and $H_{yy}=0.5$, respectively. We find that the $h_{xy}$ functions ($h_{xy}^{\theta=0}$, $h_{xy}^{\theta=0.5}$, $h_{xy}^{\theta=1}$ and $h_{xy}^{\rm{DFA}}$) for the four algorithms are independent of $\rho$. This finding is very important since it distinguishes the temporal cross-correlations quantified by MF-DCCA algorithms and the cross-sectional correlation quantified by $\rho$. Two uncorrelated time series may exhibit long-term power-law cross-correlation. In addition, the centered MF-X-DMA and the MF-X-DFA algorithms give similarly very accurate estimates of the scaling exponents with $h_{xy}^{\theta=0.5} \approx h_{xy}^{\rm{DFA}}$, $H_{xy}=0.8$ for the up panel and $H_{xy}=0.3$ for the low panel. In contrast, Fig.~\ref{Fig:MFDCCA:bFBM}(c) shows that the backward and forward MF-X-DMA algorithms perform slightly worse and $h_{xy}^{\theta=0} \approx h_{xy}^{\theta=1}$.

In order to compare the performance of the four algorithms, we calculate the difference between the estimated exponent $h_{xy}$ and the theoretical exponent $H_{xy}$:
\begin{equation}
 \Delta{h}_{xy} = h_{xy}-H_{xy}.
 \label{Eq:MFDCCA:dH}
\end{equation}
Figure~\ref{Fig:MFDCCA:bFBM}(e) shows the dependence of $\Delta{h}_{xy}$ with respect to $H_{yy}$ with a fixed $H_{xx}=0.1$ for $\rho=0.5$ (top panel), $\rho=0.3$ (middle panel) and $\rho=0.1$ (bottom panel), while Fig.~\ref{Fig:MFDCCA:bFBM}(f) shows the dependence of $\Delta{h}_{xy}$ with respect to $H_{xx}=H_{yy}$ for $\rho=0.9$ (top panel), $\rho=0.5$ (middle panel) and $\rho=0.1$ (bottom panel). All the $\Delta{h}_{xy}$ values in the two plots are less than 0.01, implying that all the four algorithms give good estimates. It is interesting to observe that $h_{xy}^{\theta=0} \approx h_{xy}^{\theta=1}$ for all the cases. In addition, the centered MF-X-DMA and the MF-X-DFA algorithms outperform the backward and forward MF-X-DMA algorithms. We note that these conclusions also hold for other bFBMs. The relative performance between the centered MF-X-DMA and the MF-X-DFA algorithms are a little bit complicated. When $H_{xx}\neq H_{yy}$, as shown in Fig.~\ref{Fig:MFDCCA:bFBM}(e), the two algorithms have comparable performance since $\Delta h_{xy}^{\theta=0.5}\approx \Delta h_{xy}^{\rm{DFA}}\approx0$. When $H_{xx} = H_{yy}$, as shown in Fig.~\ref{Fig:MFDCCA:bFBM}(f), the centered MF-X-DMA algorithm slightly outperforms the MF-X-DFA algorithm. In summary, the centered MF-X-DMA algorithm ($\theta=0.5$) is recommended for analyzing bivariate fractional Brownian motions.

\begin{figure*}[htb]
\centering
\includegraphics[width=5.5cm]{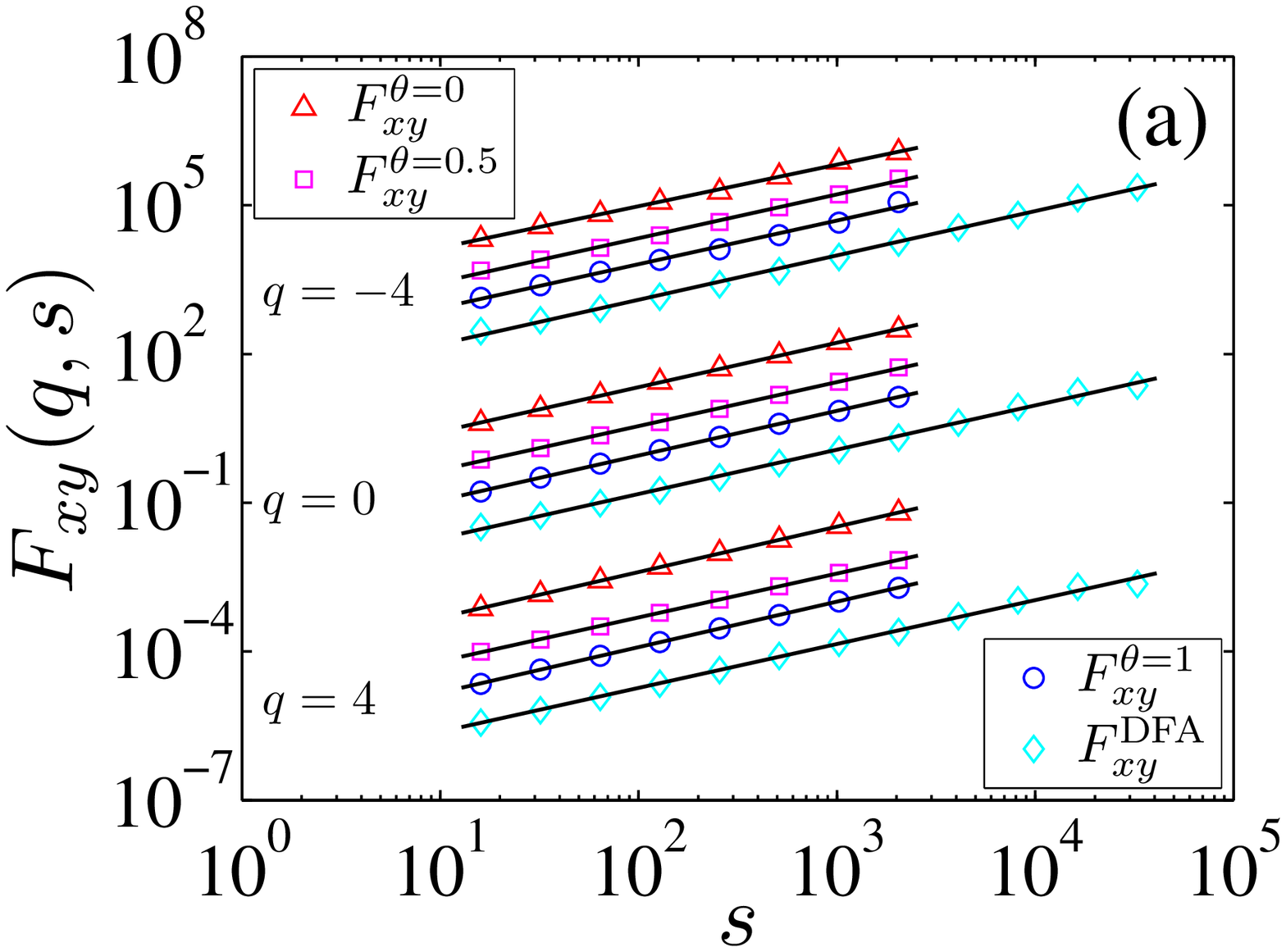}
\includegraphics[width=5.5cm]{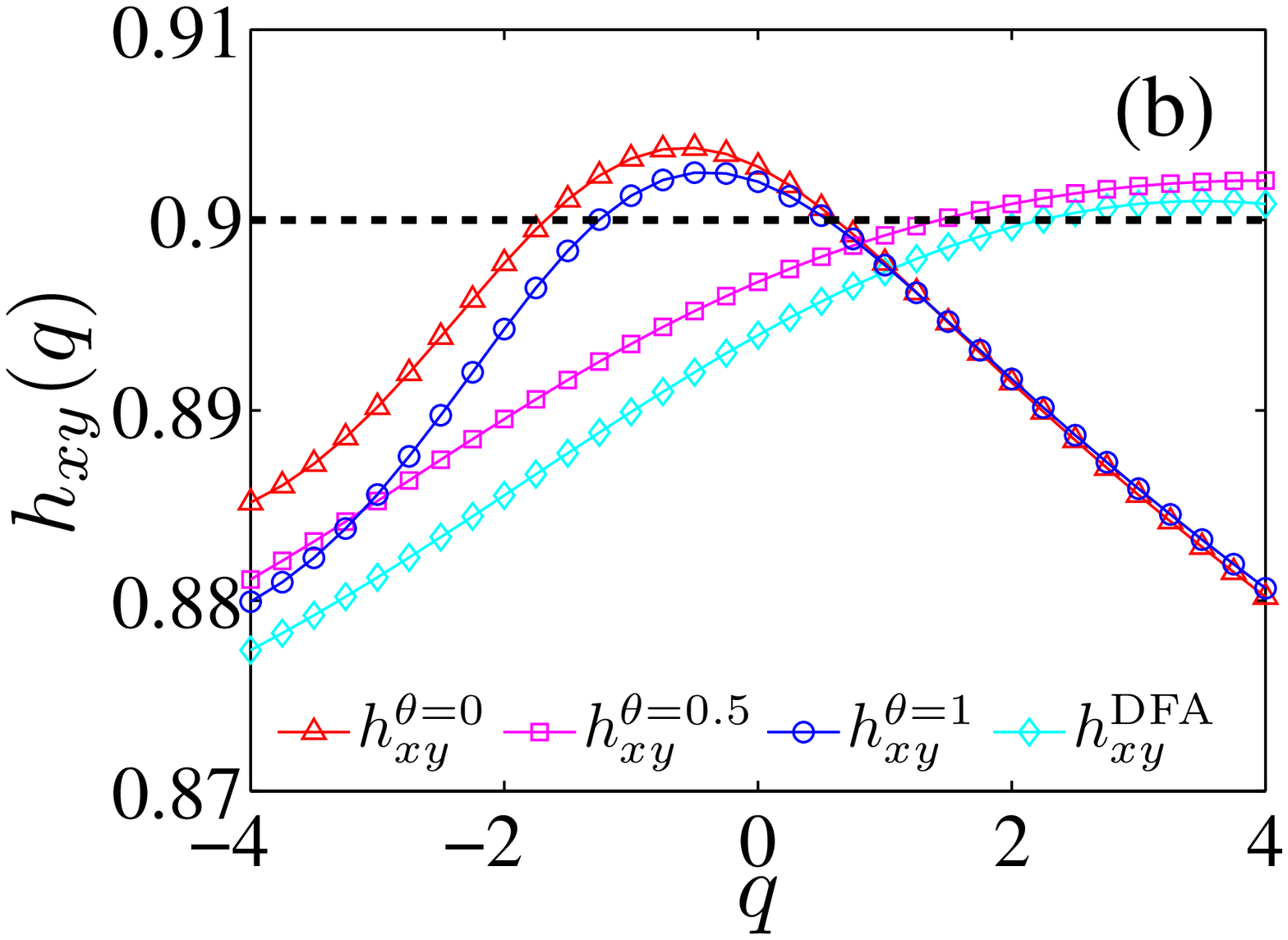}
\includegraphics[width=5.5cm]{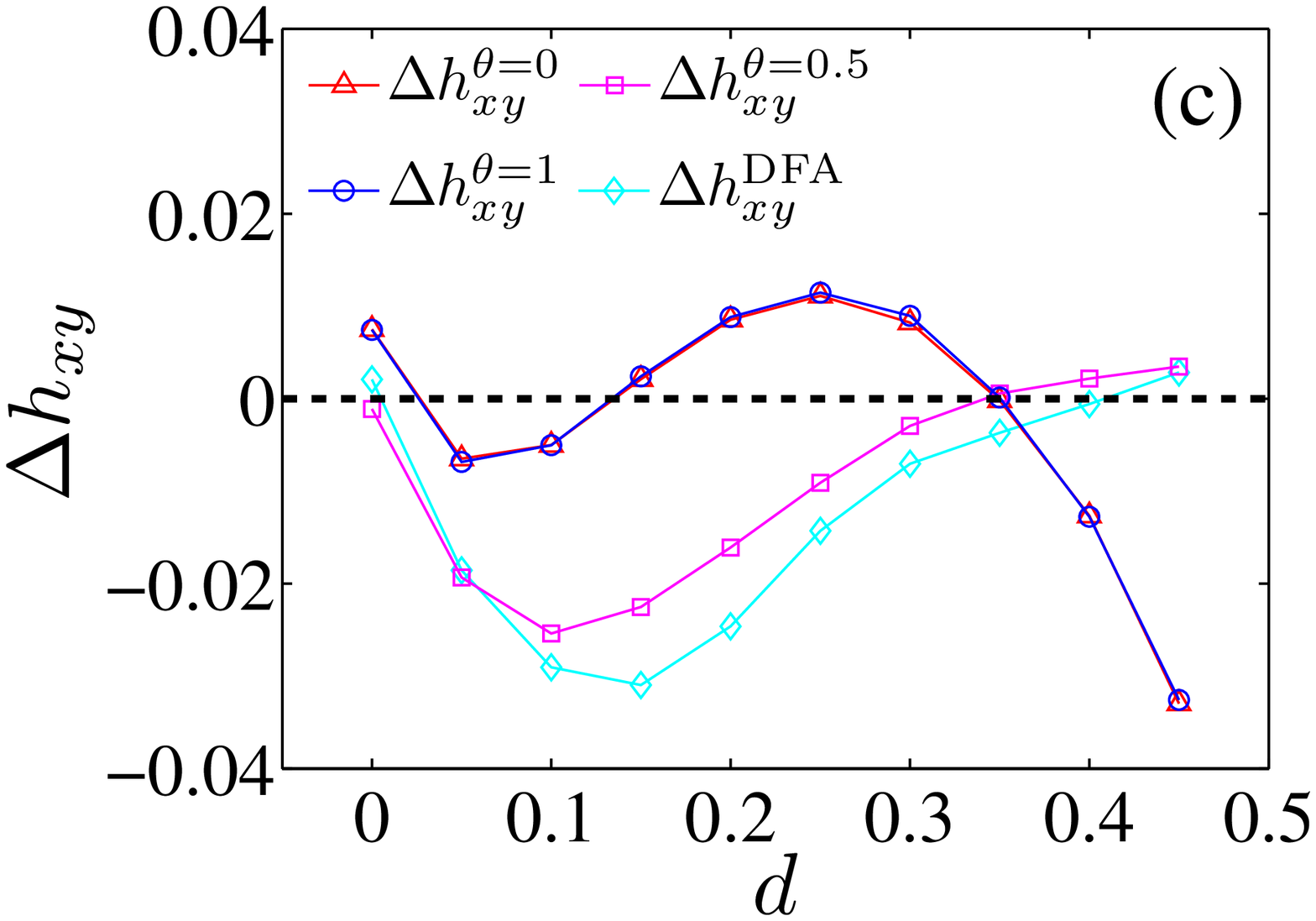}
\includegraphics[width=5.5cm]{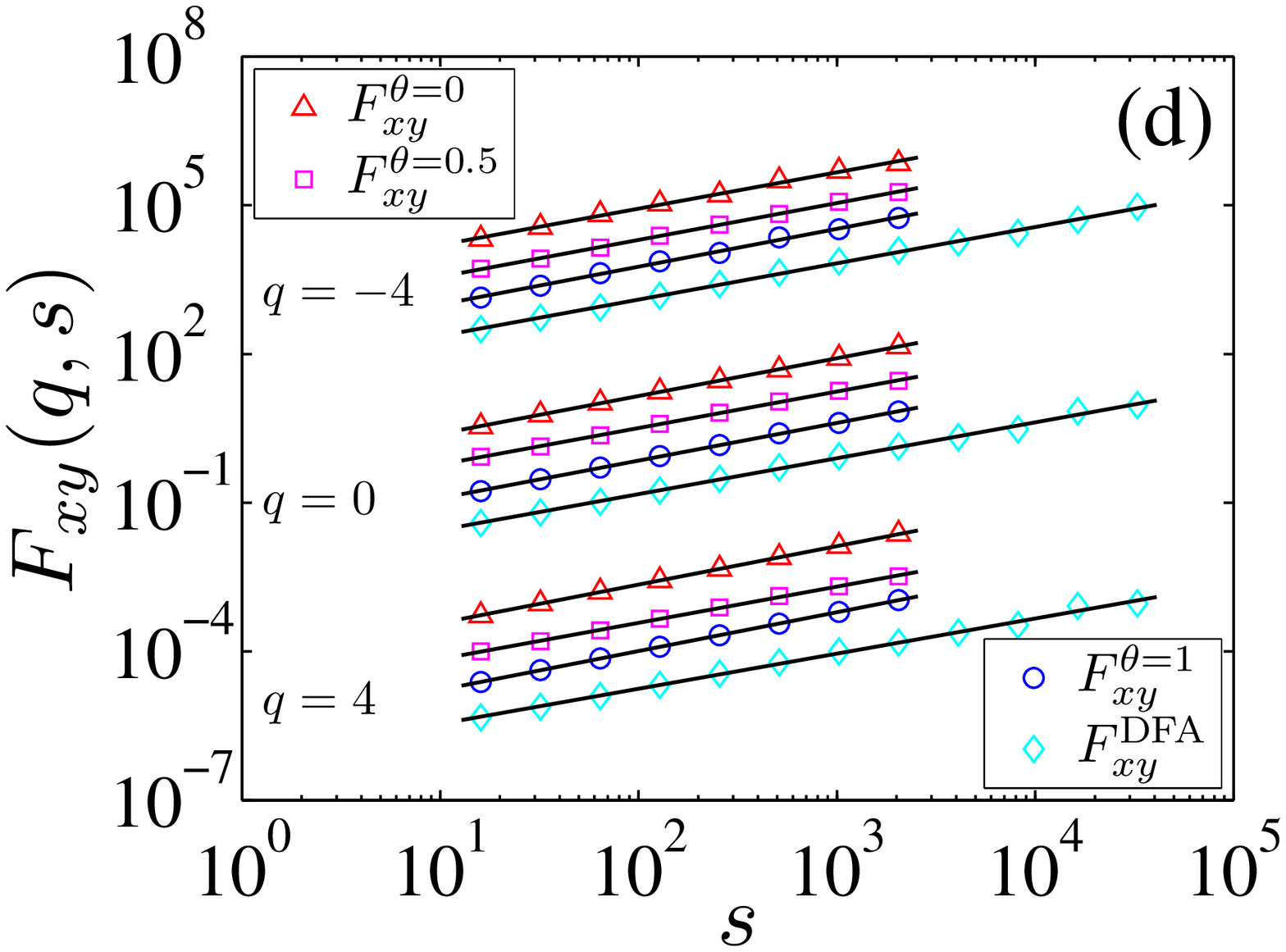}
\includegraphics[width=5.5cm]{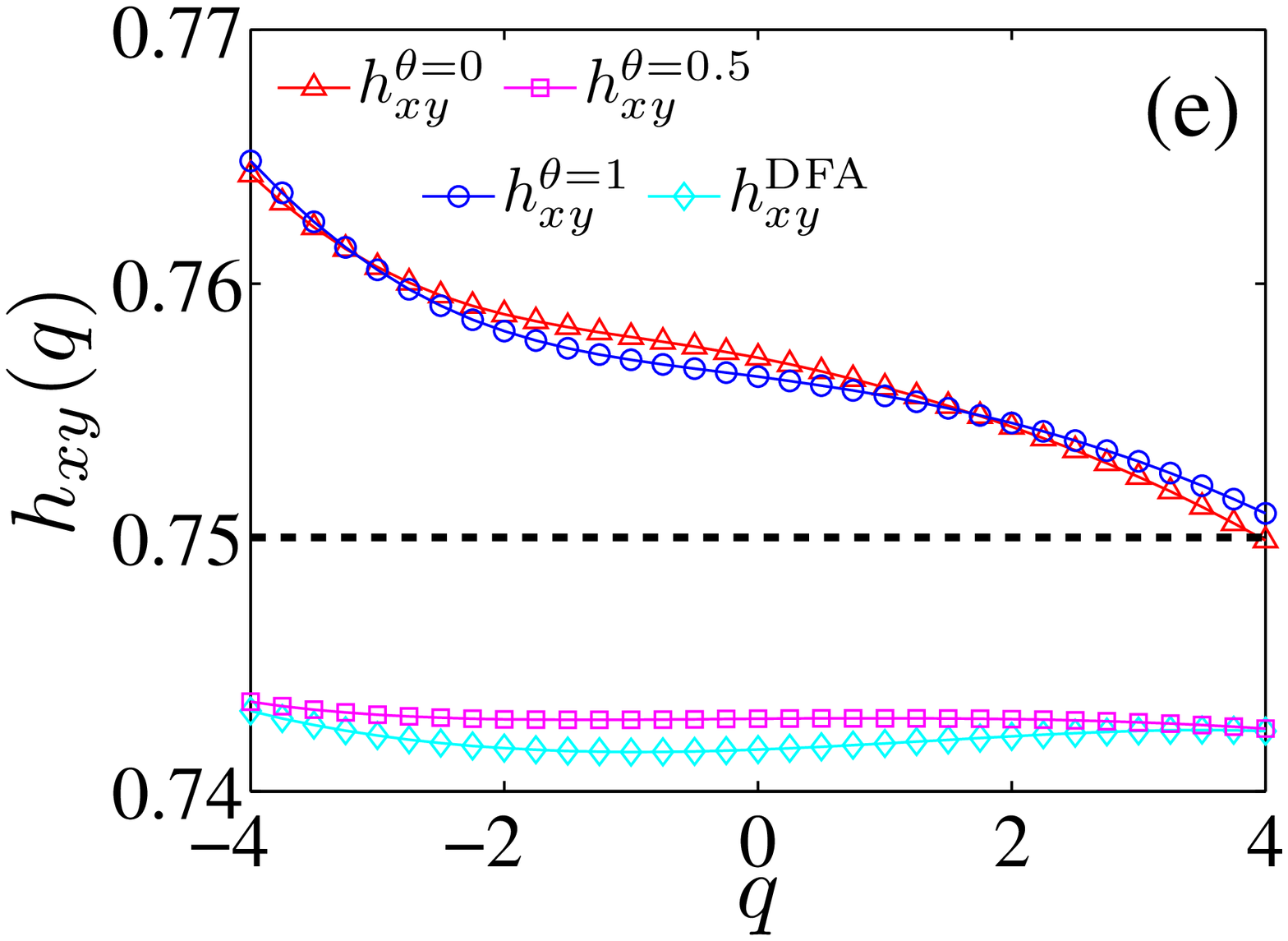}
\includegraphics[width=5.5cm]{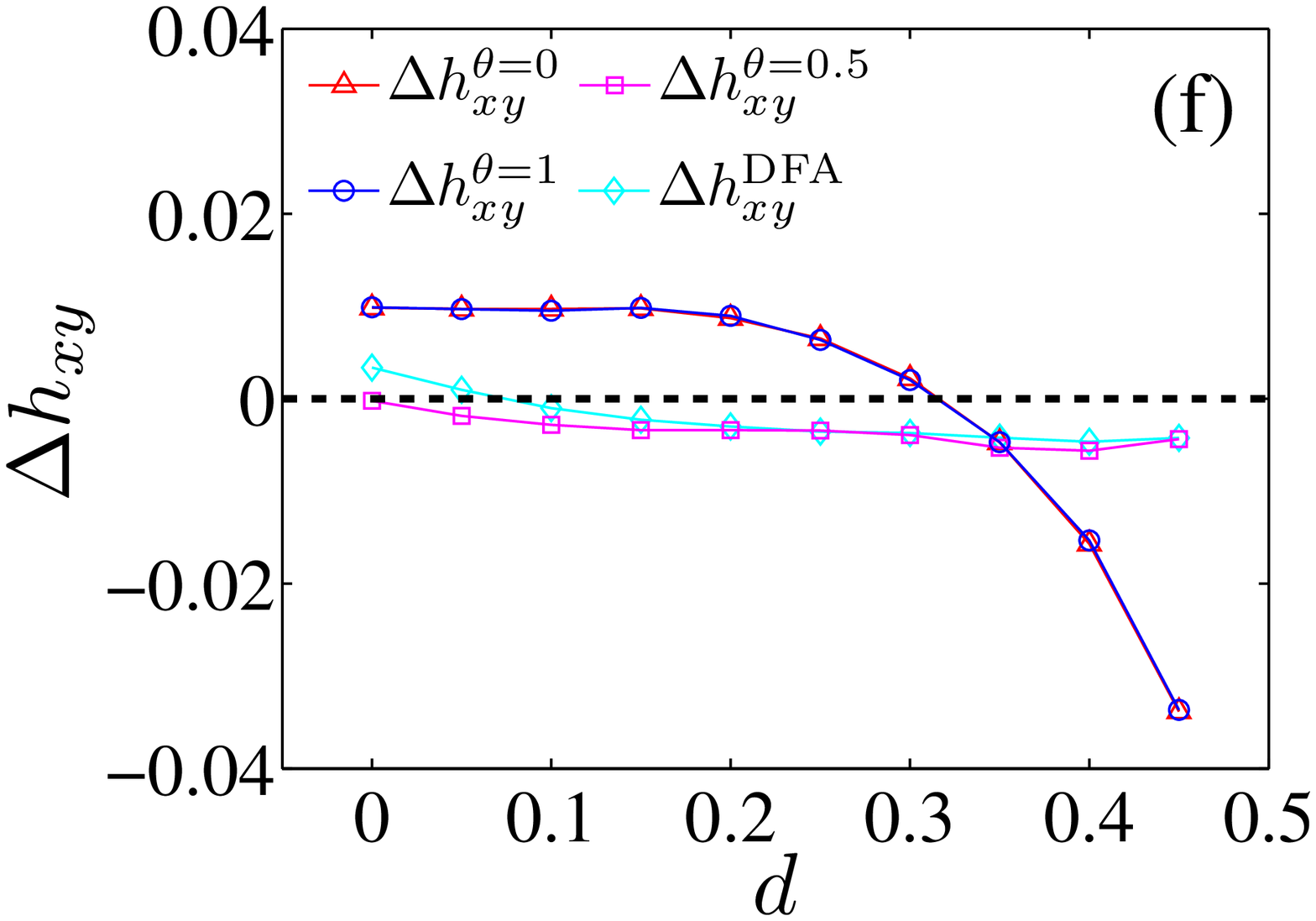}
\caption{\label{Fig:MFDCCA:ARFIMA} (Color online) Multifractal detrended cross-correlation analysis of two-component ARFIMA processes. Comparisons are performed among three MF-X-DMA algorithms with $\theta=0$, 0.5 and 1 and the MF-X-DFA method. (a) Power-law dependence of the fluctuation functions $F_{xy}(q,s)$ with respect to the scale $s$ for $q=-4$, $q=0$, and $q=4$ for the process in Eq.~(\ref{Eq:ARFIMA}) with $d_1=d_2=0.4$. The straight lines are the best power-law fits to the data. The results have been translated vertically for better visibility. (b) Scaling exponents $h_{xy}(q)$ for the process in Eq.~(\ref{Eq:ARFIMA}) with $d_1=d_2=0.4$. (c) Differences $\Delta{h}_{xy}$ between the estimated scaling exponents $h_{xy}$ and the theoretical exponents $H_{xy}$ for the process in Eq.~(\ref{Eq:ARFIMA}) with different $d$ values where $d_1=d_2=d$. (d) Power-law dependence of the fluctuation functions $F_{xy}(q,s)$ with respect to the scale $s$ for $q=-4$, $q=0$, and $q=4$ for the process in Eq.~(\ref{Eq:ARFIMA:I}) with $d_1=0.1$ and $d_2=0.4$. (e) Scaling exponents $h_{xy}(q)$ for the process in Eq.~(\ref{Eq:ARFIMA:I}) with $d_1=0.1$ and $d_2=0.4$. (f) Differences $\Delta{h}_{xy}$ between $h_{xy}$ and $H_{xy}$ for the process in Eq.~(\ref{Eq:ARFIMA:I}) with different $d$ values where $d_1=d_2=d$.}
\end{figure*}

\subsection{Two-component ARFIMA processes}

The power-law auto-correlations in stochastic variables can be modeled by an ARFIMA process \cite{Hosking-1981-Bm}:
\begin{equation}
  z(t)=Z(d,t)+\epsilon(t),
  \label{Eq:ARFIMA:z}
\end{equation}
where $d\in(0,0.5)$ is a memory parameter, $\epsilon_z$ is an independent and identically distributed Gaussian variable, and
\begin{equation}
  Z(d,t)=\sum_{n=1}^{\infty}a_n(d)z(t-n),
  \label{Eq:ARFIMA:Zdt}
\end{equation}
in which $a_n(d)$ is the weight
\begin{equation}
  a_n(d)=d\Gamma(n-d)/[\Gamma(1-d)\Gamma(n+1)].
  \label{Eq:ARFIMA:an}
\end{equation}
The Hurst index $H_{zz}$ is related to the memory parameter $d$ by \cite{Podobnik-Ivanov-Biljakovic-Horvatic-Stanley-Grosse-2005-PRE,Podobnik-Stanley-2008-PRL}
\begin{equation}
 H_{zz}=0.5+d.
 \label{Eq:H:d}
\end{equation}
For the two-component ARFIMA processes discussed below, we take $Z=X$ or $Y$.

The two-component ARFIMA process is defined as follows \cite{Podobnik-Horvatic-Ng-Stanley-Ivanov-2008-PA}:
\begin{equation}
    \left\{
    \begin{array}{ll}
        x(t)= WX(d_1,t) + (1-W) Y(d_2,t)+\epsilon_x(t)\\
        y(t)=(1-W)X(d_1,t) + W Y(d_2,t)+\epsilon_y(t)
    \end{array}
    \right.,
\label{Eq:ARFIMA}
\end{equation}
where $W\in[0.5,1]$ quantifies the coupling strength between the two processes $x(t)$ and $y(t)$. When $W=1$, $x(t)$ and $y(t)$ are fully decoupled and become two separate ARFIMA processes as defined in Eq.~(\ref{Eq:ARFIMA:z}). The cross-correlation between $x(t)$ and $y(t)$ increases when $W$ decreases from 1 to 0.5 \cite{Podobnik-Horvatic-Ng-Stanley-Ivanov-2008-PA}. To our knowledge, no general expression has been analytically derived for $H_{xy}$. When $d_1>d_2$, the Hurst index $H_{xx}$ of $x(t)$ decreases from $0.5+d_1$ to certain value greater than $0.5+d_2$ when $W$ decreases from 1 to 0.5 \cite{Podobnik-Horvatic-Ng-Stanley-Ivanov-2008-PA}. In other words, $H_{xx}$ locates within the interval $[0.5+d_2,0.5+d_1]$. When $d_1=d_2=d$, {\textit{i.e.}} $d_1\to d_2$, we obtain that
\begin{equation}
  H_{xx}=H_{yy}=0.5+d,
  \label{Eq:Eq:ARFIMA:Hxx:d1d2}
\end{equation}
which does not depend on the value of $W$.

When $W=1$ and $\epsilon_x(t)=\epsilon_y(t)=\epsilon(t)$, the two-component ARFIMA process becomes \cite{Podobnik-Stanley-2008-PRL}
\begin{equation}
    \left\{
    \begin{array}{ll}
        x(t)= X(d_1,t) +\epsilon(t)\\
        y(t)= Y(d_2,t) +\epsilon(t)
    \end{array}
    \right..
\label{Eq:ARFIMA:I}
\end{equation}
If $x$ and $y$ are long-range power-law cross-correlated, it has been analytically derived that Eq.~(\ref{Eq:Hxy:Hxx:Hyy}) holds \cite{Podobnik-Grosse-Horvatic-Ilic-Ivanov-Stanley-2009-EPJB}.

The top panel (a-c) of Fig.~\ref{Fig:MFDCCA:ARFIMA} shows the results for the process in Eq.~(\ref{Eq:ARFIMA}). Figure~\ref{Fig:MFDCCA:ARFIMA}(a) illustrates in log-log scale the dependence of the fluctuation functions $F_{xy}(q,s)$ with respect to the scale $s$ for $q=-4$, $q=0$, and $q=4$ for the process in Eq.~(\ref{Eq:ARFIMA}) with $d_1=d_2=0.4$. Nice power-law relations are observed, which are also evident for other $(d_1,d_2)$ pairs. Figure \ref{Fig:MFDCCA:ARFIMA}(b) shows the corresponding scaling exponents $h_{xy}(q)$ for the process in Eq.~(\ref{Eq:ARFIMA}) with $d_1=d_2=0.4$. We note that the equation $h_{xy}(q)=[h_{xx}(q)+h_{yy}(q)]/2$ holds for all the four curves. For the four algorithms, $h_{xy}(q)$ is close to the horizontal line $H=0.9$, indicating that all the four algorithms correctly unveil the fractal nature of the two-component ARFIMA process. For $q=2$, this plot shows that the MF-X-DFA gives the best estimate of $h_{xy}$. Figure \ref{Fig:MFDCCA:ARFIMA}(c) depicts the differences $\Delta{h}_{xy}$ between $h_{xy}$ and $H_{xy}$ with $q=2$ for the process in Eq.~(\ref{Eq:ARFIMA}) with different $d$ values where $d_1=d_2=d$. It is found that, (1) the two MF-X-DMA algorithms with $\theta=0$ and $\theta=1$ have the same performance, (2) the two MF-X-DMA algorithms with $\theta=0$ and $\theta=1$ perform better than the MF-X-DFA and the MF-X-DMA with $\theta=0.5$ for relatively small $d$ values, and (3) the two MF-X-DMA algorithms with $\theta=0$ and $\theta=1$ perform worse for large $d$ values.

The bottom panel (d-f) of Fig.~\ref{Fig:MFDCCA:ARFIMA} shows the results for the process in Eq.~(\ref{Eq:ARFIMA:I}). Figure \ref{Fig:MFDCCA:ARFIMA}(d) illustrates in log-log scale the dependence of the fluctuation functions $F_{xy}(q,s)$ with respect to the scale $s$ for $q=-4$, $q=0$, and $q=4$ for the process in Eq.~(\ref{Eq:ARFIMA:I}) with $d_1=0.1$ and $d_2=0.4$. Nice power-law relations are observed, which are also evident for other $(d_1,d_2)$ pairs. Figure \ref{Fig:MFDCCA:ARFIMA}(e) shows the corresponding scaling exponents $h_{xy}(q)$ for the process in Eq.~(\ref{Eq:ARFIMA:I}) with $d_1=0.1$ and $d_2=0.4$. Again, the equation $h_{xy}(q)=[h_{xx}(q)+h_{yy}(q)]/2$ holds for all the four curves. For the four algorithms, $h_{xy}(q)$ is close to the horizontal line $H=0.75$, indicating that all the four algorithms correctly unveil the fractal nature of the two-component ARFIMA process. For $q=2$, this plot shows that the MF-X-DMA algorithms with $\theta=0$ and $\theta=1$ give the best estimate of $h_{xy}$. Figure \ref{Fig:MFDCCA:ARFIMA}(f) shows the differences between $h_{xy}$ and $H_{xy}$ with $q=2$ for the process in Eq.~(\ref{Eq:ARFIMA:I}) with different $d$ values where $d_1=d_2=d$. It is found that the MF-X-DFA and the MF-X-DMA with $\theta=0.5$ outperform the other two algorithms and give comparably nice estimates.

He and Chen have investigated the two-component ARFIMA process defined in Eq.~(\ref{Eq:ARFIMA}) of different lengths utilizing the DCCA method (the MF-X-DFA method with $q=2$) and the DMCA method (the MF-X-DMA method with $\theta=0$ and $q=2$) and found that the DMCA method performs better in most cases and performs worse in a few cases \cite{He-Chen-2011b-PA}. However, both methods are prone to underestimate the exponents $h_{xy}$ \cite{He-Chen-2011b-PA}. Our results shown in Fig.~\ref{Fig:MFDCCA:ARFIMA}(c) are consistent with their results for $d=0.15$, 0.25 and 0.35 in the sense that the MF-X-DMA method with $\theta=0$ outperforms the MF-X-DFA method. However, we have obtained better estimates for $h_{xy}$ and there is no systematic underestimation. For instance, the three MF-X-DMA methods give $h_{xy}\approx0.85$ or $\Delta{h}_{xy}\approx0$ when $d=0.35$ as shown in Fig.~\ref{Fig:MFDCCA:ARFIMA}(c).

There are two subtle issues that might worsen the estimation of $h_{xy}$. As stated by Podobnik and Stanley \cite{Podobnik-Stanley-2008-PRL}, they introduced a cutoff length $M=10^4$ in their numerical simulations and let the sum run from 1 to $M$, i.e., they set $a_j=0$ for $j>M$. Our numerical experiments show that this cutoff seems optimal and a smaller or larger cutoff will worsen the estimation of the exponents. This finding applies both the MF-X-DMA and the MF-X-DFA. In addition, we stress that the upper bound of the scaling range for the MF-X-DMA algorithms should not be too large, because each moving averages are calculated within a window of size $s$. Let us take the MF-X-DMA algorithm with $\theta=0$ as an example. In this case, the moving averages of the first $s-1$ data points are not well defined. The bias becomes more significant for large window size $s$. When $s$ is large, the $F_{xy}(q,s)$ function bends downwards and the overall slope flattens. Similar arguments apply for other MF-X-DMA algorithms with different $\theta$ values. In contrast, the MF-X-DFA algorithm does not suffer from this finite-size effect.

\begin{figure*}[htb]
\centering
\includegraphics[width=5.5cm]{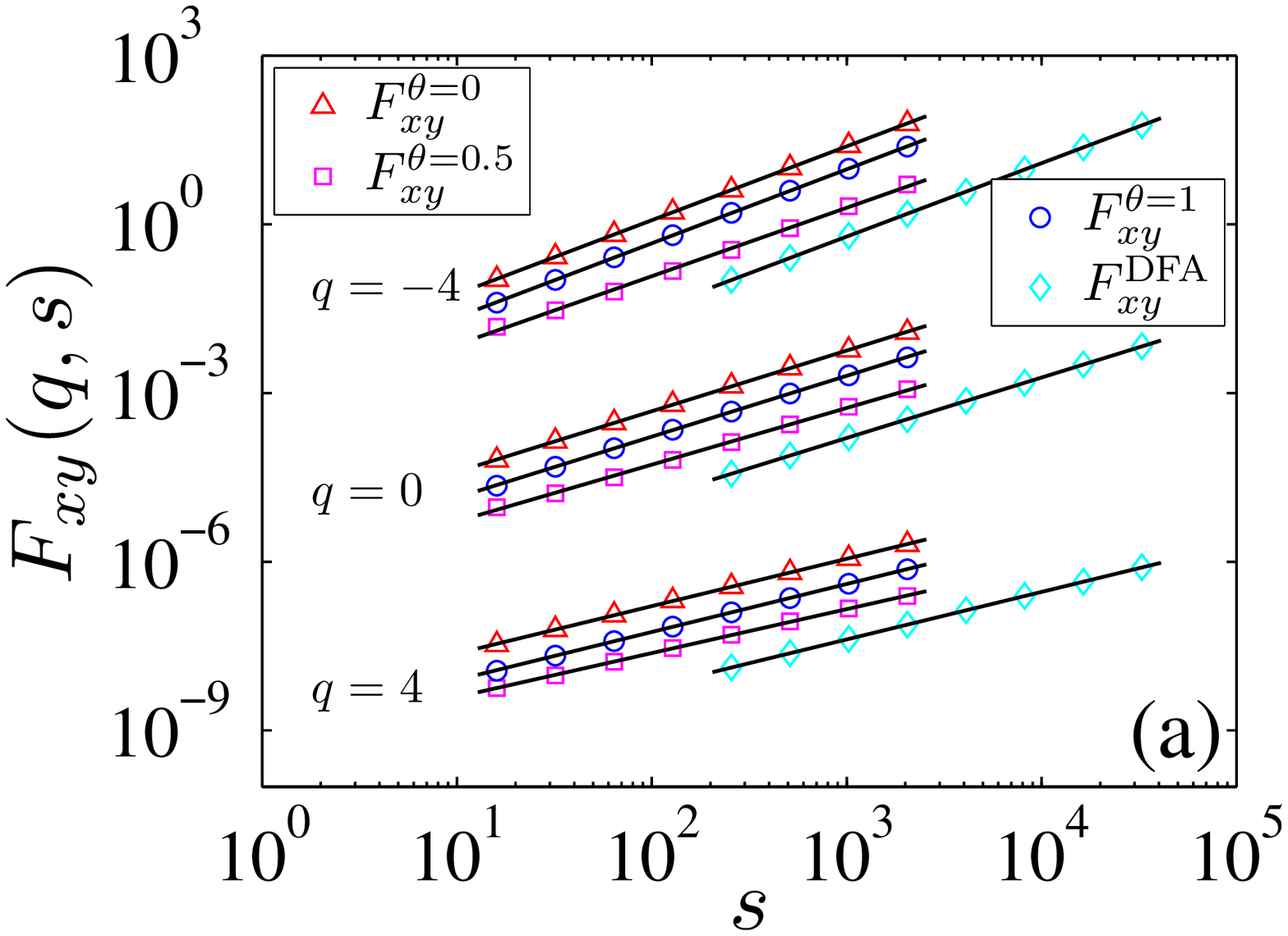}
\includegraphics[width=5.5cm]{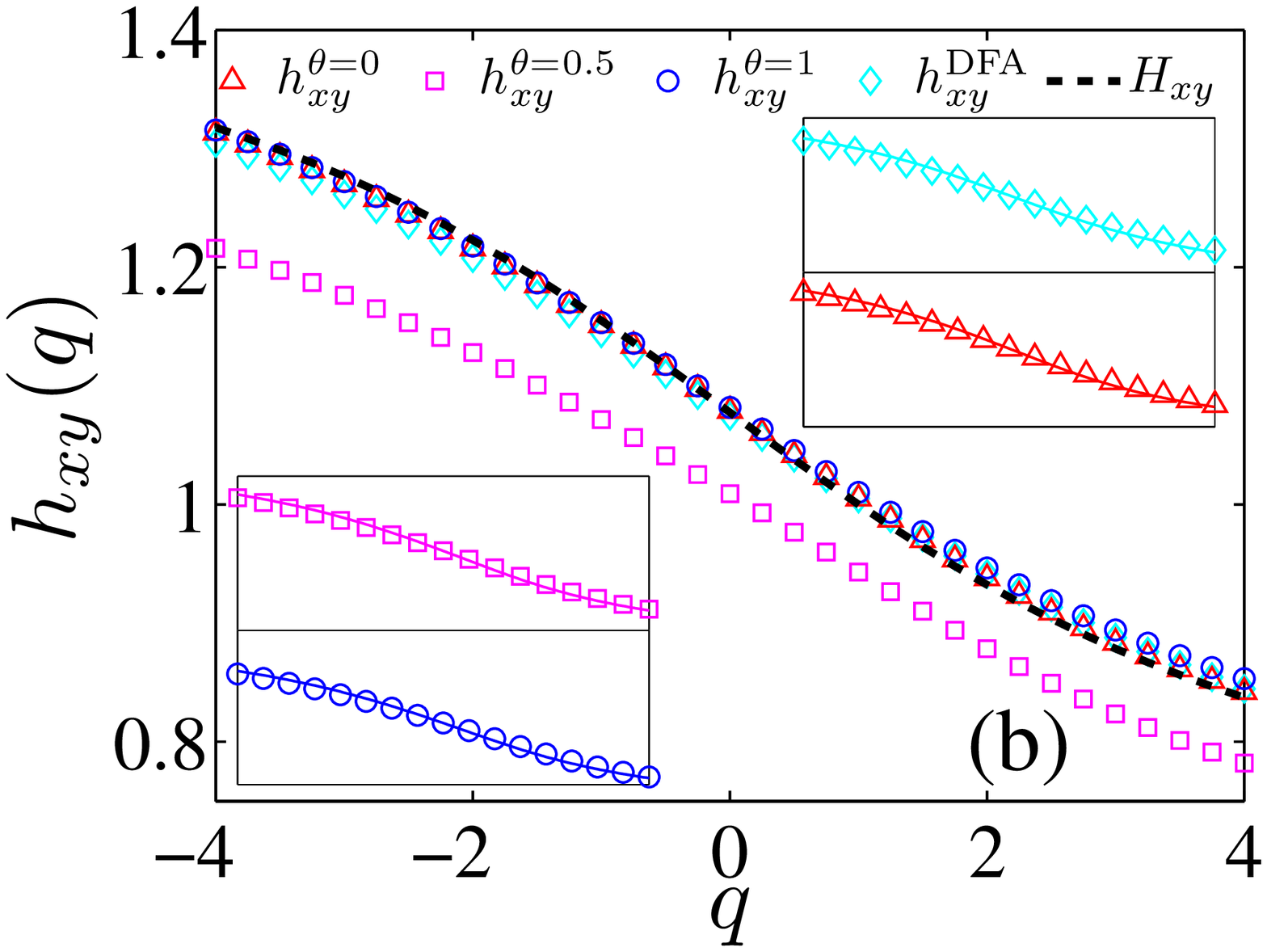}
\includegraphics[width=5.5cm]{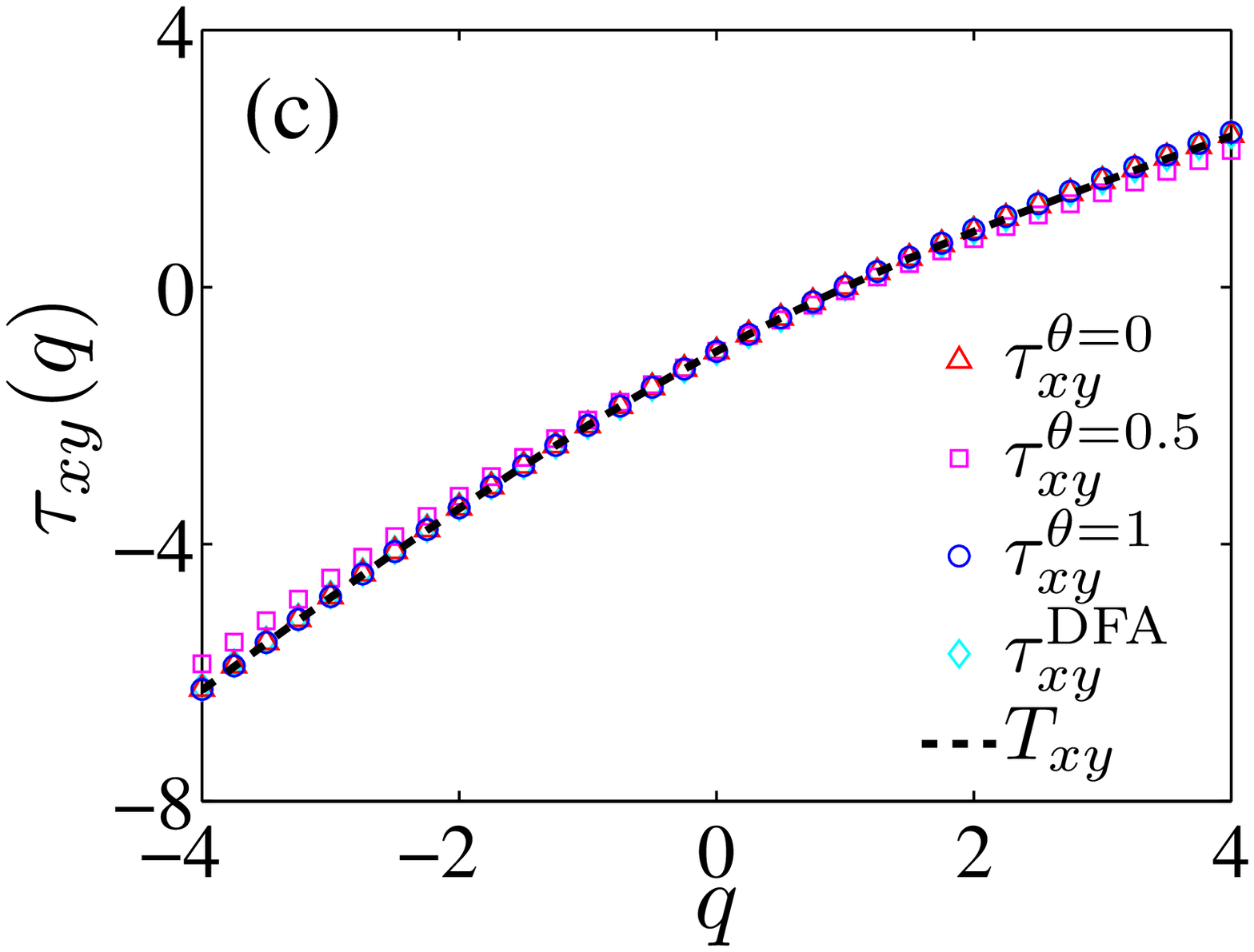}
\includegraphics[width=5.5cm]{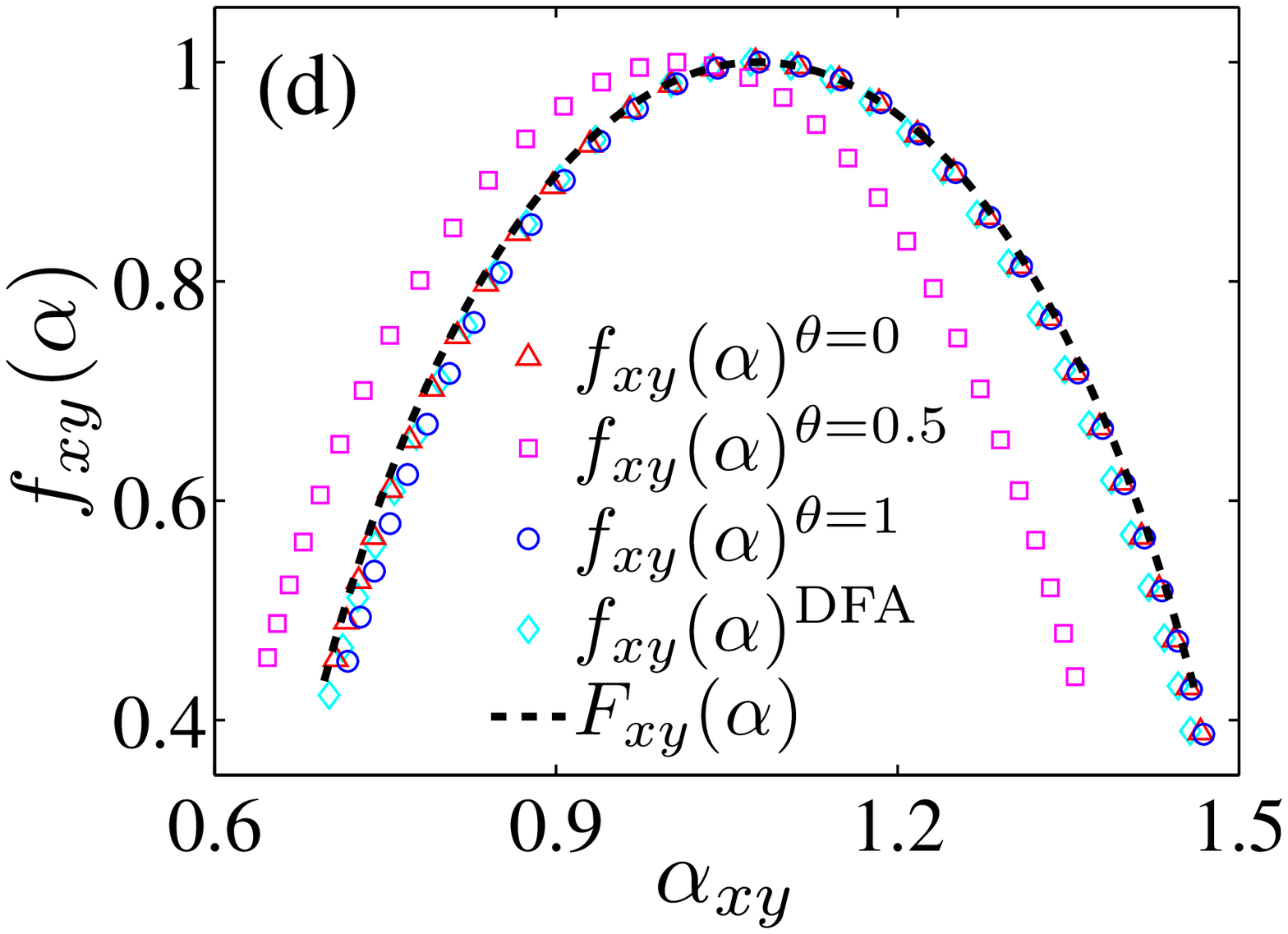}
\includegraphics[width=5.5cm]{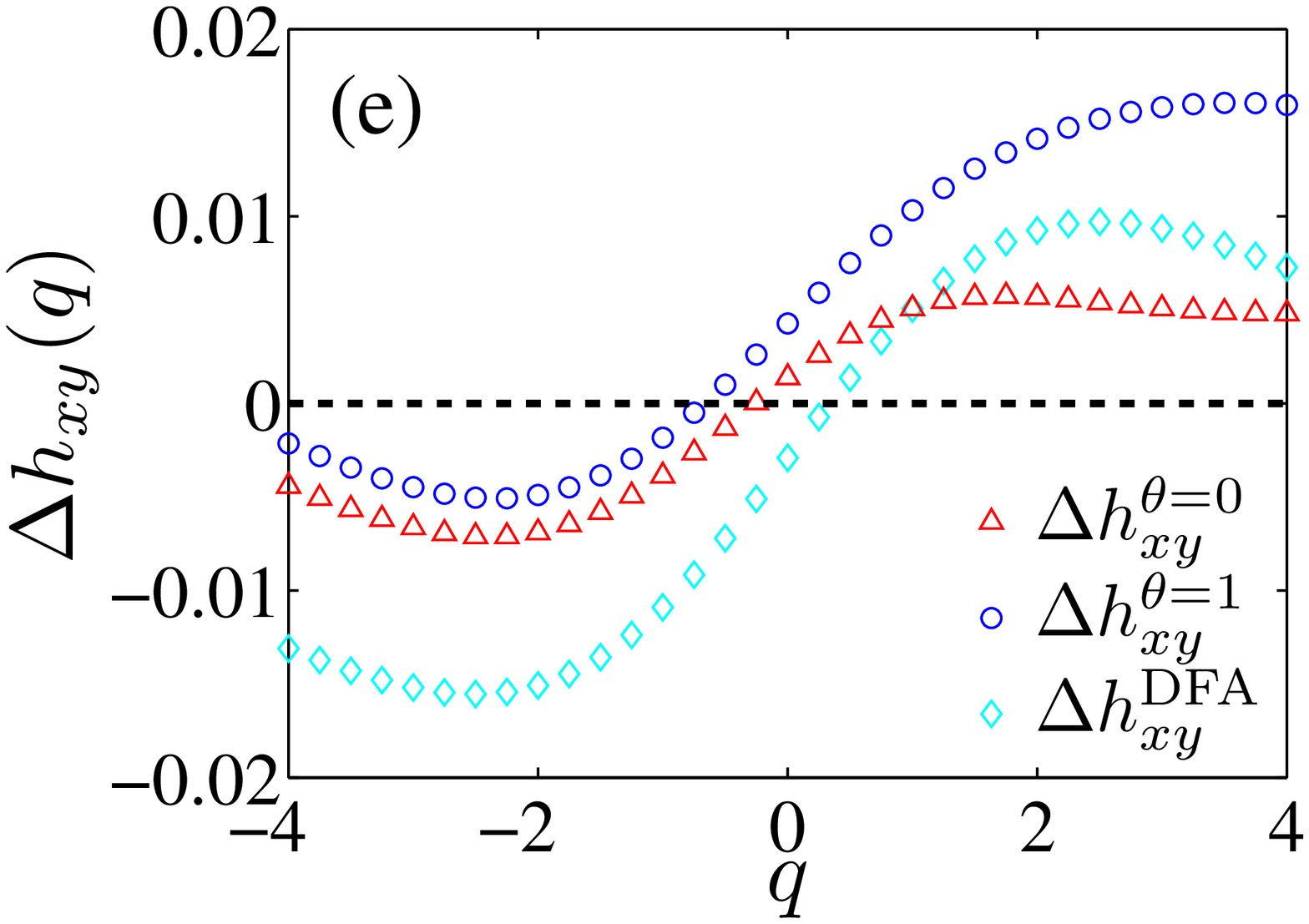}
\includegraphics[width=5.5cm]{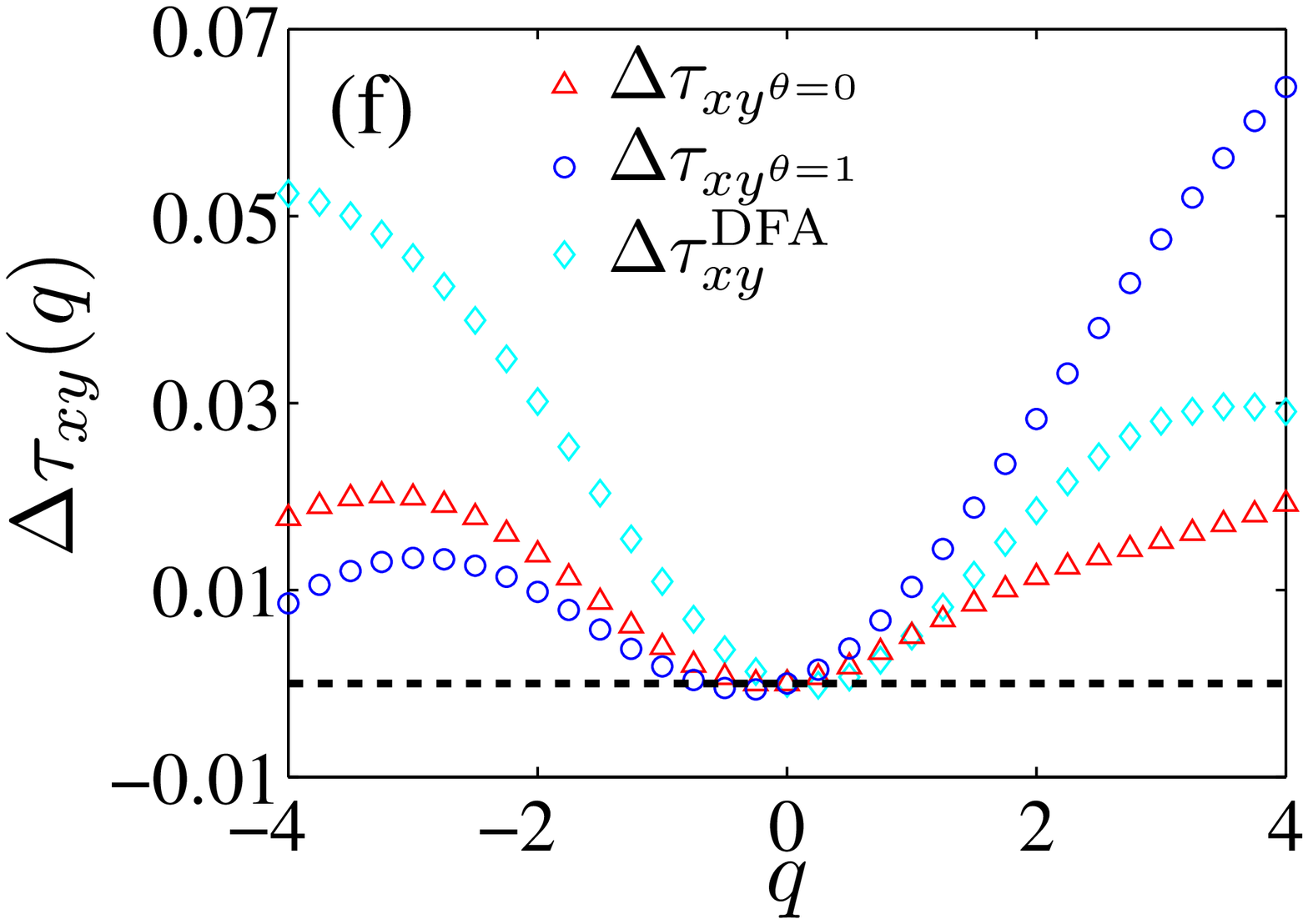}
\caption{\label{Fig:MFDCCA:pModel} (Color online) Multifractal detrended cross-correlation analysis of two cross-correlated binomial measures generated from the $p$-model with $p_x=0.3$ and $p_y=0.4$. Comparisons are performed among three MF-X-DMA algorithms with $\theta=0$, 0.5 and 1 and the MF-X-DFA method. (a) Power-law dependence of the fluctuation functions $F_{xy}(q,s)$ with respect to the scale $s$ for $q=-4$, $q=0$, and $q=4$. The straight lines are the best power-law fits to the data. The results have been translated vertically for better visibility. (b) Scaling exponents $h_{xy}(q)$ with the theoretical values as a dashed line. The insets show the $h_{xy}(q)$ curves and the corresponding $[h_{xx}(q)+h_{yy}(q)]/2$ curves, verifying the relation $h_{xy}(q)=[h_{xx}(q)+h_{yy}(q)]/2$. (c) Multifractal mass exponents $\tau(q)$ obtained from the MFDMA and MFDFA methods with the theoretical curve shown as a dashed line. (d) Multifractal spectra $f(\alpha)$ with respect to the singularity strength $\alpha$ for the four methods. The dashed curve is the theoretical multifractal spectrum. (e) Differences $\Delta{h}_{xy}(q)$ between the estimated mass exponents and their theoretical values for the three algorithms: MF-X-DFA, MF-X-DMA with $\theta=0$ and MF-X-DMA with $\theta=1$. (f) Differences $\Delta{\tau}(q)$ between the estimated mass exponents and their theoretical values for the three algorithms: MF-X-DFA, MF-X-DMA with $\theta=0$ and MF-X-DMA with $\theta=1$.}
\end{figure*}

\subsection{Multifractal binomial measures}

We construct two binomial measures $\{x(i): i = 1, 2, \cdots, 2^k\}$ and $\{y(i): i = 1, 2, \cdots, 2^k\}$ from the $p$-model with known analytic multifractal properties as a first example \cite{Meneveau-Sreenivasan-1987-PRL}. Each multifractal signal is obtained in an iterative way. We start with the zeroth iteration $k = 0$, where the data set $z(i)$ consists of one value, $z^{(0)}(1)= 1 $. In the $k$th iteration, the data set $\{z^{(k)}(i): i = 1, 2, \cdots, 2^k\}$ is obtained from
\begin{equation}
  \begin{array}{l}
    z^{(k)}(2i-1)= p_z z^{(k-1)}(i)\\
    z^{(k)}(2i)  = (1-p_z)z^{(k-1)}(i)
\end{array}
  \label{Eq:pModel}
\end{equation}
for $i = 1, 2, \cdots, 2^{k-1}$. We notice that there are typos in the formula in Ref.~\cite{Zhou-2008-PRE}. When $k\to\infty$, $z^{(k)}(i)$ approaches to a binomial measure, whose scaling exponent function $H_{zz}(q)$ has an analytic form \cite{Halsey-Jensen-Kadanoff-Procaccia-Shraiman-1986-PRA,Meneveau-Sreenivasan-1987-PRL}
\begin{equation}
 H_{zz}(q) = 1/q-\log_2[p_z^q+(1-p_z)^q]/q.
 \label{Eq:pModel:Hzz}
\end{equation}
According to Eq.~(\ref{Eq:MFDCCA:tau}), we have
\begin{equation}
 {\cal{T}}_{zz}(q) = -\log_2[p_z^q+(1-p_z)^q].
 \label{Eq:pModel:Tzz}
\end{equation}

In our simulation, we have performed $k=16$ iterations with $p_x =0.3$ for $x(i)$ and $p_y = 0.4$ for $y(i)$. The analytic scaling exponent functions $H_{xx}(q)$ and $H_{yy}(q)$ of $x$ and $y$ are expressed in Eq.~(\ref{Eq:pModel:Hzz}). The two time series $x$ and $y$ are strongly correlated with a coefficient of 0.82, which is originated from the fact that the two sequences are constructed according to the same rules. The results are depicted in Fig.~\ref{Fig:MFDCCA:pModel}.

Figure \ref{Fig:MFDCCA:pModel}(a) illustrates the power-law dependence of $F_{xy}(q,s)$ against $s$ for the four algorithms. Since the time series is not very long, we investigate $-4\leq{q}\leq4$ to ensure the convergence of the $q$th moments \cite{Lvov-Podivilov-Pomyalove-Procaccia-Vandembroucq-1998-PRE,Zhou-Sornette-Yuan-2006-PD}. For the MF-X-DMA algorithms, there is a finite-size effect since the moving averages at the ends of the time series are ill-defined. This effect becomes significant that deteriorates the estimation of $F_{xy}$ for large scales $s$. The scaling range is chosen as $[2^4, 2^{11}]$ for the three MF-X-DMA methods. In contrast, the MF-X-DFA method performs bad if the same scaling range is adopted. We use $[2^8,2^{15}]$ for the MF-X-DFA method which seems optimal. The algorithm-specific selection of the scaling range reveals the difference in the applicability of the two types of methods. Figure \ref{Fig:MFDCCA:pModel}(a) shows that the power-law scaling is excellent for both positive and negative $q$ values.

The power-law scaling exponents ($h_{xy}^{\rm{DFA}}$, $h_{xy}^{\theta=0}$, $h_{xy}^{\theta=0.5}$, and $h_{xy}^{\theta=1}$) are presented in Fig.~\ref{Fig:MFDCCA:pModel}(b), while the mass scaling exponents ($\tau_{xy}^{\rm{DFA}}$, $\tau_{xy}^{\theta=0}$, $\tau_{xy}^{\theta=0.5}$ and $\tau_{xy}^{\theta=1}$) and the multifractal spectra ($f_{xy}^{\rm{DFA}}$, $f_{xy}^{\theta=0}$, $f_{xy}^{\theta=0.5}$ and $f_{xy}^{\theta=1}$) are illustrated in Fig.~\ref{Fig:MFDCCA:pModel}(c) and Fig.~\ref{Fig:MFDCCA:pModel}(d). It is evident that the MF-X-DMA method with $\theta=0.5$ fails in a large part to correctly estimate the exponents, while the other three methods work much better. This finding is consistent with the conclusion that the MF-DMA method with $\theta=0.5$ performs much worse than the MF-DFA method and the MF-DMA methods with $\theta=0$ and $\theta=1$ \cite{Gu-Zhou-2010-PRE}.

The insets of Fig.~\ref{Fig:MFDCCA:pModel}(b) show an interesting feature for all the four algorithms that
\begin{equation}
 h_{xy}(q) = [h_{xx}(q)+h_{yy}(q)]/2,
 \label{Eq:MFDCCA:pModel:hhh}
\end{equation}
no matter how accurate the estimates of an algorithm is. Similar relationship holds for individual monofractal ARFIMA signals \cite{Podobnik-Stanley-2008-PRL} and individual binomial measures \cite{Gu-Zhou-2010-PRE}. Hence, we can give the ``theoretical'' expression of $H_{xy}(q)$ as follows
\begin{equation}
 H_{xy}(q) = [H_{xx}(q)+H_{yy}(q)]/2,
 \label{Eq:MFDCCA:pModel:HHH}
\end{equation}
where $H_{xx}(q)$ and $H_{yy}(q)$ are given in Eq.~(\ref{Eq:pModel:Hzz}). The theoretical line is plotted in Fig.~\ref{Fig:MFDCCA:pModel}(b) as a dashed line. According to Eq.~(\ref{Eq:MFDCCA:tau}), we obtain
\begin{equation}
 {\cal{T}}_{xy}(q) = [{\cal{T}}_{xx}(q)+{\cal{T}}_{yy}(q)]/2,
 \label{Eq:MFDCCA:pModel:TTT}
\end{equation}
which is shown in Fig.~\ref{Fig:MFDCCA:pModel}(c) for comparison. Similarly, we have
\begin{equation}
 {\cal{F}}_{xy}(\alpha) = [{\cal{F}}_{xx}(\alpha)+{\cal{F}}_{yy}(\alpha)]/2,
 \label{Eq:MFDCCA:pModel:FFF}
\end{equation}
which is illustrated in Fig.~\ref{Fig:MFDCCA:pModel}(d) as a dashed curve.

In order to further assess the performance of the MF-X-DFA method and the two MF-X-DMA methods with $\theta=0$ and $\theta=1$, we compare the empirical estimates of $h_{xy}(q)$ and $\tau_{xy}(q)$ with the theoretical values of $H_{xy}(q)$ and ${\cal{T}}_{xy}(q)$ by calculating the relative errors:
\begin{equation}
 \Delta{h}_{xy}(q) = h_{xy}(q)-H_{xy}(q)
 \label{Eq:MFDCCA:pModel:dH}
\end{equation}
and
\begin{equation}
 \Delta\tau_{xy}(q) = \tau_{xy}(q)-{\cal{T}}_{xy}(q).
 \label{Eq:MFDCCA:pModel:dTau}
\end{equation}
which are shown in Fig.~\ref{Fig:MFDCCA:pModel}(e) and Fig.~\ref{Fig:MFDCCA:pModel}(f). Roughly speaking, the MF-X-DMA algorithm with $\theta=1$ performs best and the MF-DFA algorithm performs worst for most negative $q$ values and the MF-X-DMA method with $\theta=0$ performs best and the MF-X-DMA method with $\theta=1$ performs worst for most negative $q$ values. On average, the backward MF-X-DMA method ($\theta=0$) has the best performance and is thus recommended.

\begin{figure*}[htb]
\centering
\includegraphics[width=5.5cm]{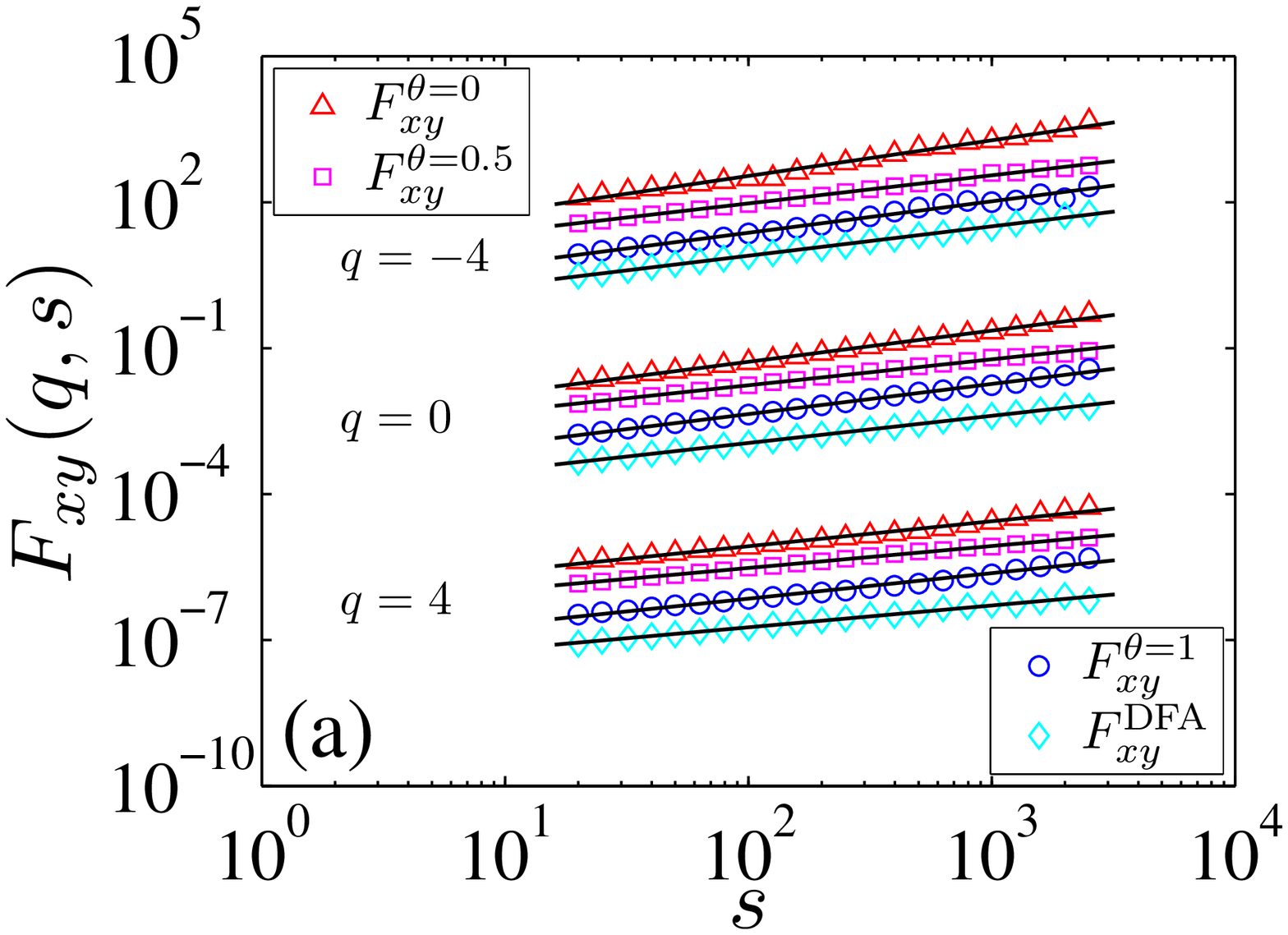}
\includegraphics[width=5.5cm]{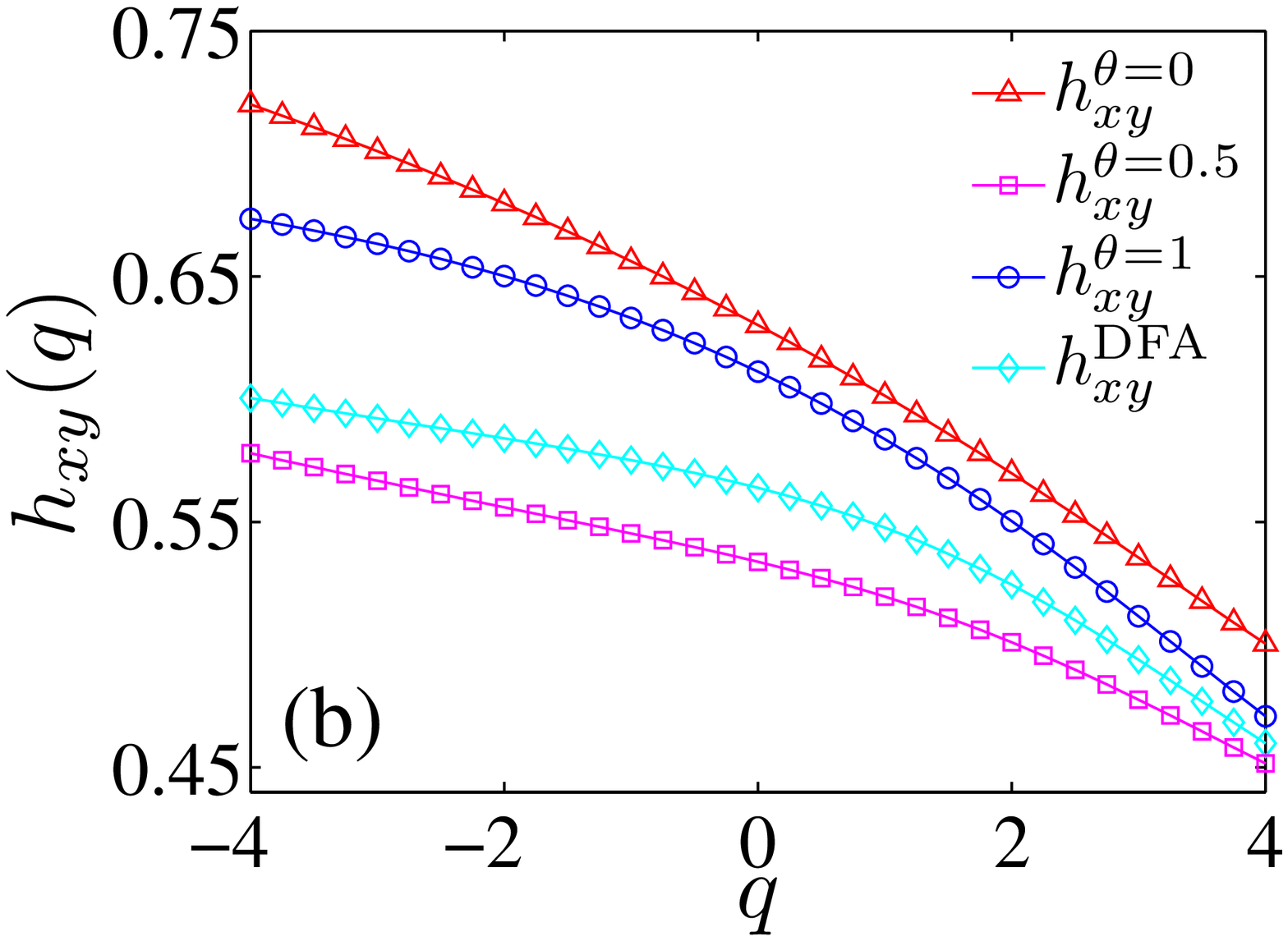}
\includegraphics[width=5.5cm]{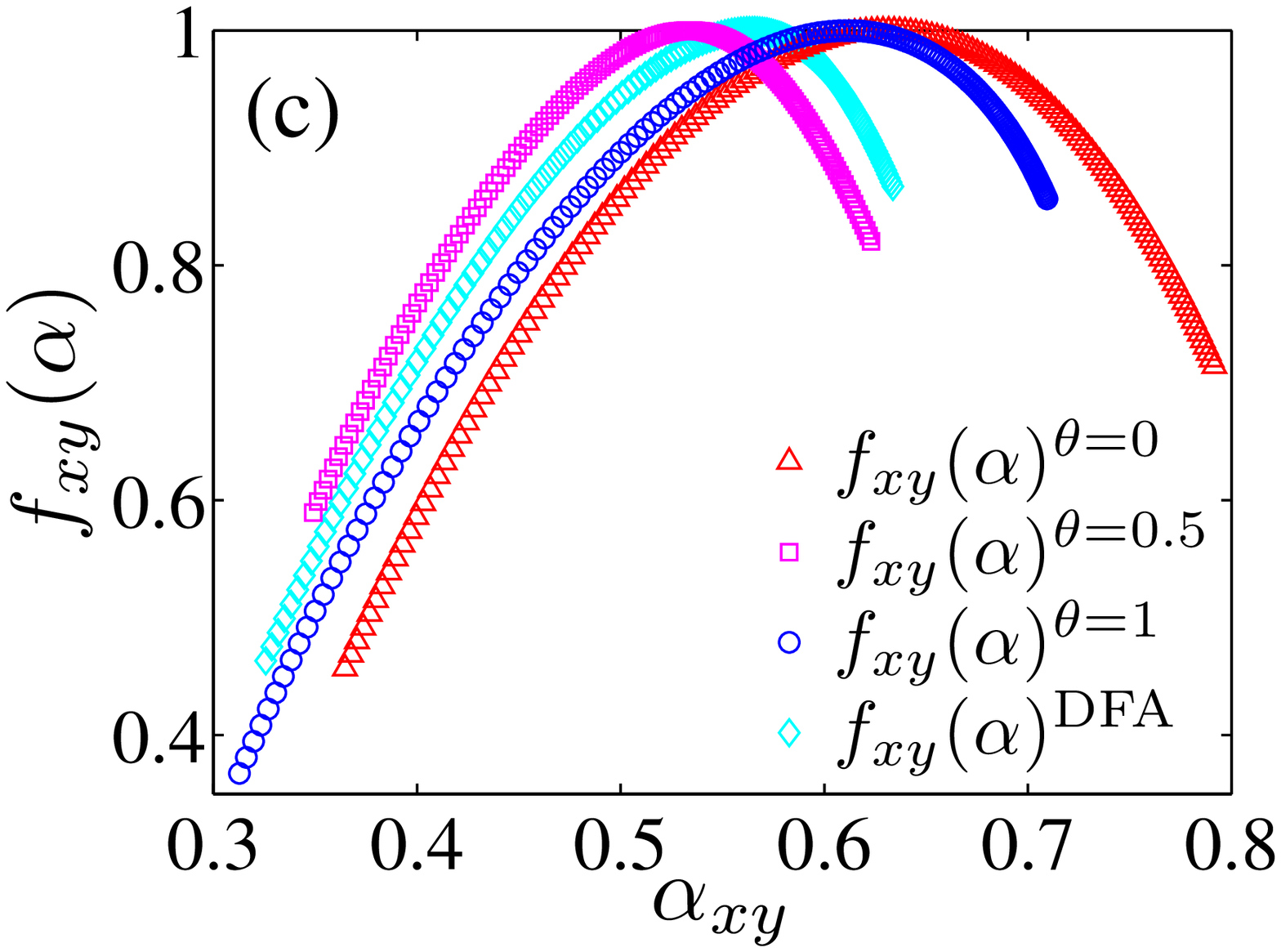}
\includegraphics[width=5.5cm]{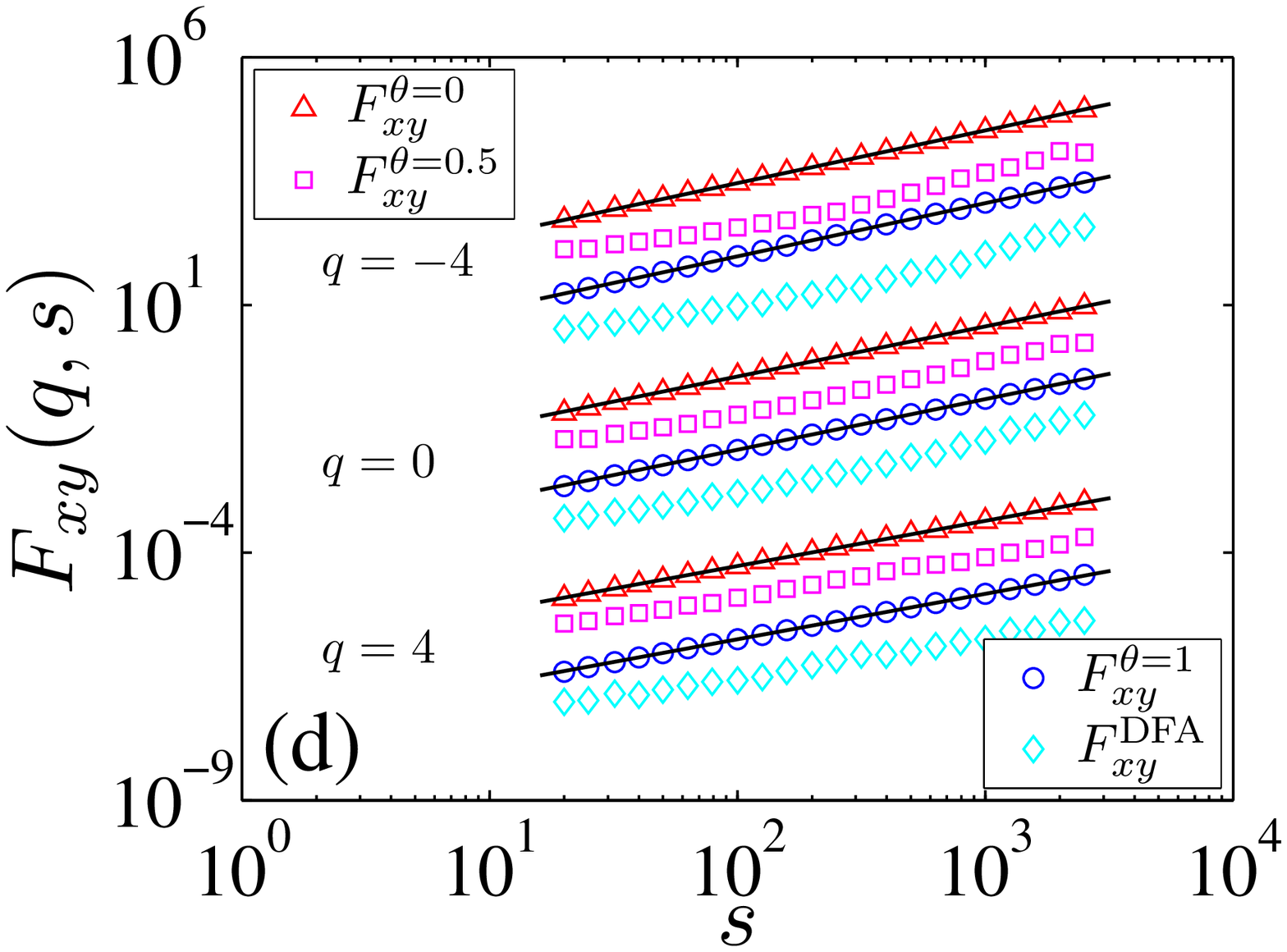}
\includegraphics[width=5.5cm]{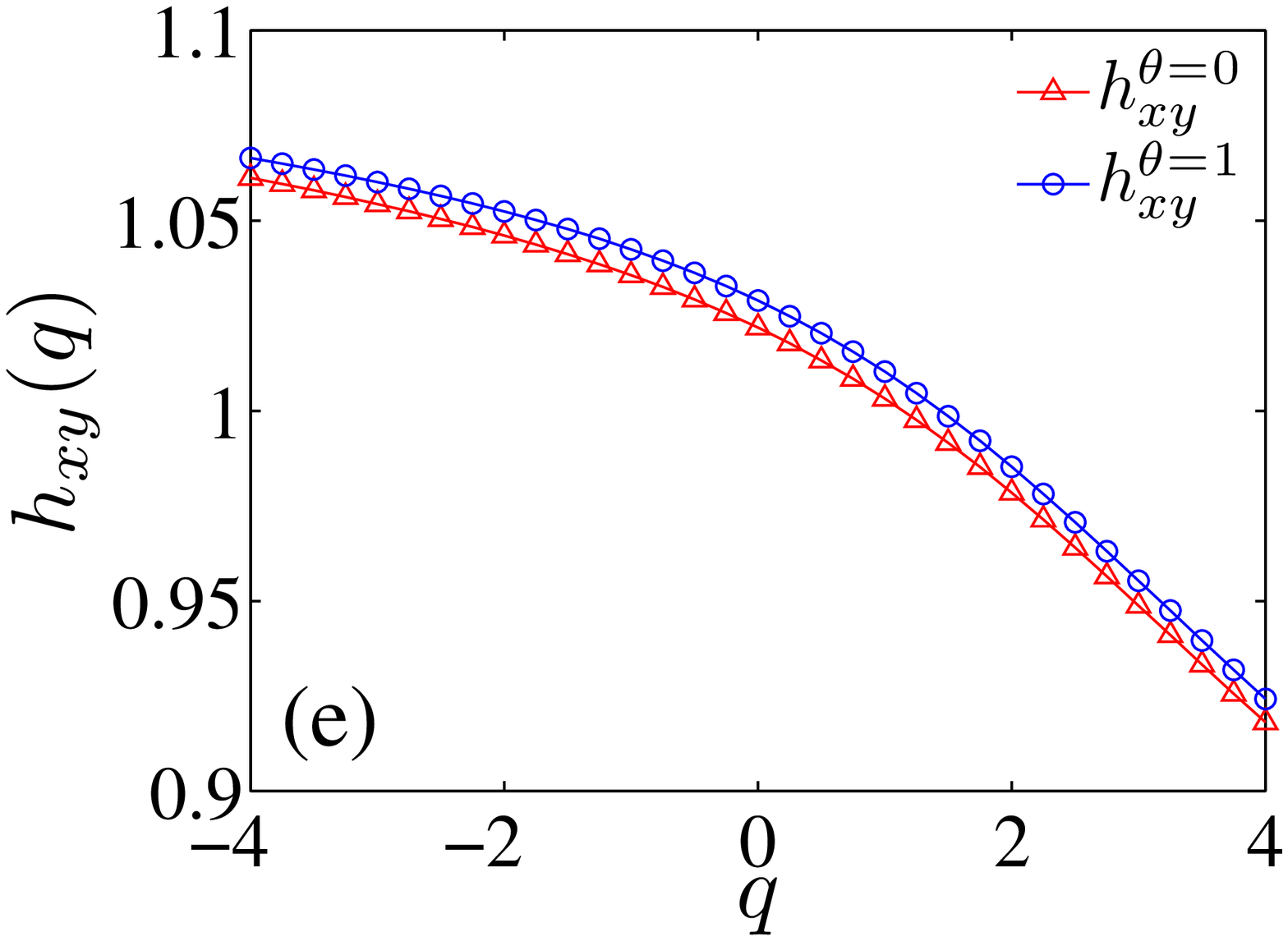}
\includegraphics[width=5.5cm]{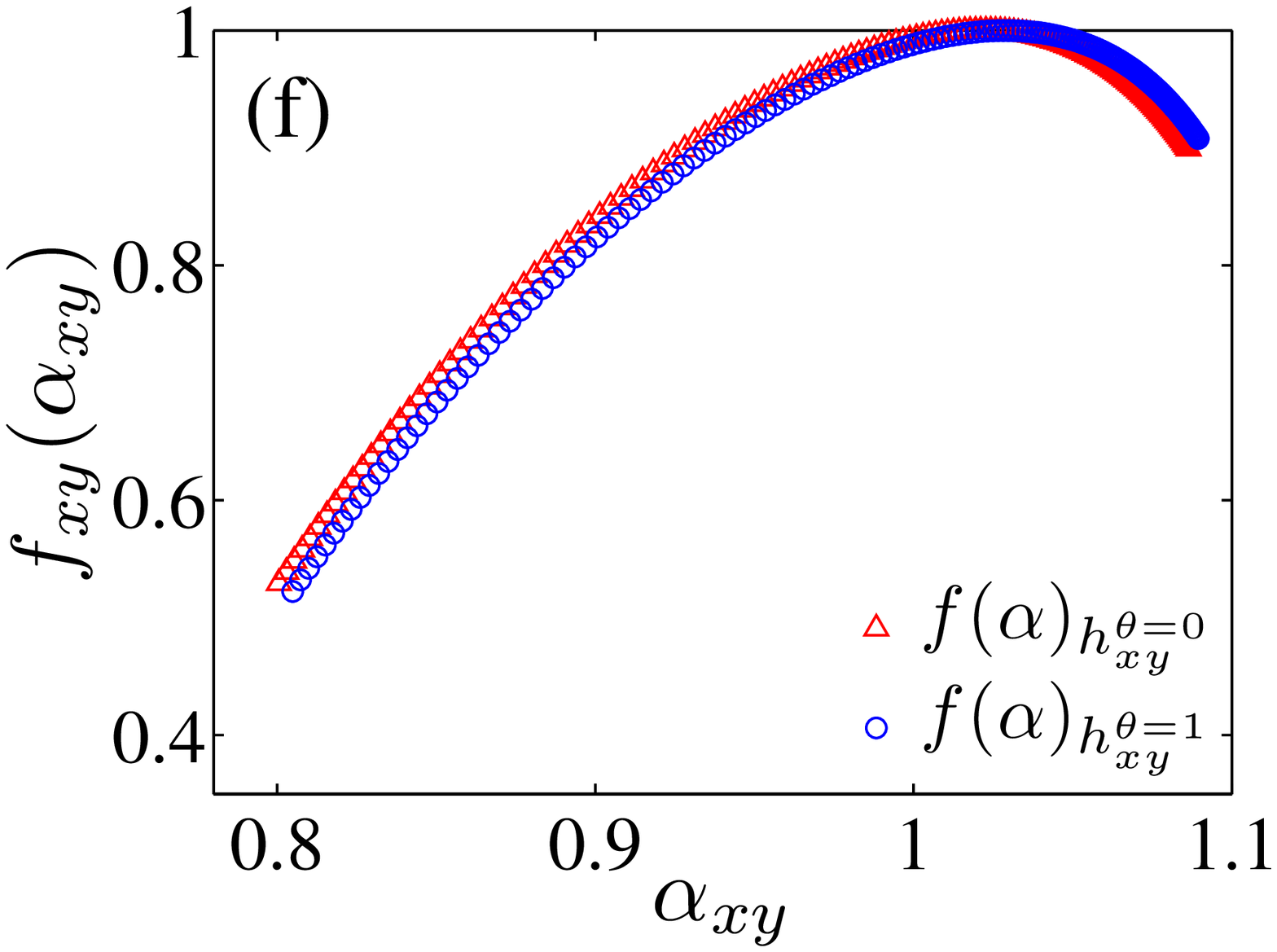}
\caption{\label{Fig:MFDCCA:Stock} (Color online) Multifractal detrended cross-correlation analysis of the return time series (a-c) and the volatility time series (d-f) for the DJIA index and the NASDAQ index. Comparisons are performed among three MF-X-DMA algorithms with $\theta=0$, 0.5 and 1 and the MF-X-DFA method. (a,d) Dependence of the fluctuation functions $F_{xy}(q,s)$ with respect to the scale $s$ for $q=-4$, $q=0$, and $q=4$. The straight lines are the best power-law fits to the data. The results have been translated vertically for better visibility. (b,e) Scaling exponents $h_{xy}(q)$ with respect to $q$. (c,f) Multifractal spectra $f_{xy}(\alpha)$ with respect to the singularity strength $\alpha$.}
\end{figure*}

\section{Application to stock market index returns and volatilities}
\label{S1:Application}

We now apply the MF-X-DMA algorithms to investigate the temporal cross-correlations of the daily return and volatility time series of the Dow Jones Industrial Average (DJIA) and the National Association of Securities Dealers Automated Quotations (NASDAQ) index. The power-law cross-correlations between the DJIA volatility and the NASDAQ volatility have been studied using the DCCA method \cite{Podobnik-Stanley-2008-PRL} and the MF-X-DFA method \cite{Zhou-2008-PRE}. The closing prices of the DJIA and the NASDAQ from 1971/2/5 to 2011/1/25 have been retrieved. The length of the time series is 10084. The return is defined as the daily difference of the logarithmic closing prices and the volatility is defined as the absolute value of the return.

Figure \ref{Fig:MFDCCA:Stock}(a) illustrates in log-log scale the dependence of the fluctuation functions $F_{xy}(q,s)$ with respect to the scale $s$ for $q=-4$, $q=0$, and $q=4$ for the returns. Excellent power laws are observed spanning over two orders of magnitude for all the four algorithms. The resulting $h_{xy}$ functions are shown in Fig.~\ref{Fig:MFDCCA:Stock}(b). All the four $h_{xy}$ functions are monotonically decreasing, indicating that the cross-correlations between the index returns exhibit multifractal nature. We also find that
\begin{equation}
 h_{xy}^{\theta=0}(q) > h_{xy}^{\theta=1}(q) > h_{xy}^{\rm{DFA}}(q) > h_{xy}^{\theta=0.5}(q)
 \label{Eq:MFDCCA:Returns}
\end{equation}
for $-4\leq q \leq 4$. When $q=2$, all the $h_{xy}$ values are less than 0.6 and particularly $h_{xy}^{\theta=0.5}\approx 0.5$. This means that there is no significant linear long-term memory in the cross-correlations of the returns. The multifractal spectra $f_{xy}(\alpha)$ are plotted in Fig.~\ref{Fig:MFDCCA:Stock}(c). The singularity widths are all significantly greater than 0, confirming that the cross-correlations possess multifractal nature.

Figure \ref{Fig:MFDCCA:Stock}(d) illustrates in log-log scale the dependence of the fluctuation functions $F_{xy}(q,s)$ with respect to the scale $s$ for $q=-4$, $q=0$, and $q=4$ for the volatilities. Excellent power-law scaling is observed in the fluctuation functions for the MF-X-DMA algorithms with $\theta=0$ and 1. However, for the MF-X-DFA algorithm and the MF-X-DMA algorithm with $\theta=0.5$, there is a crossover in each curve. If we treat each curve with two scaling ranges and perform analysis on each scaling range, the resulting $h_{xy}(q)$ functions are not monotonically decreasing and the multifractal spectra $f_{xy}(\alpha)$ are not concave. We thus focus on the two MF-X-DMA algorithms with $\theta=0$ and 1 that lead to one scaling range. The two $h_{xy}(q)$ functions are depicted in Fig.~\ref{Fig:MFDCCA:Stock}(e). The two functions are monotonically decreasing and close to each other. When $q=2$, $h_{xy}$ is close to 0.98, showing a very strong linear long-term memory in the cross-correlations between volatilities. We note that the relation $h_{xy}(q)=[h_{xx}(q)+h_{yy}(q)]/2$ does not hold, which is consistent with previous work \cite{Zhou-2008-PRE}. Figure \ref{Fig:MFDCCA:Stock}(f) plots the two multifractal spectra. The large singularity width means that the cross-correlations between the two index volatilities exhibit multifractal nature.

Our results for the volatility seems different from those in Refs.~\cite{Podobnik-Stanley-2008-PRL,Zhou-2008-PRE}. First of all, the DJIA and NASDAQ time series are much longer in the current work. More importantly, the power-law scaling in the previous works exhibits significant fluctuations \cite{Podobnik-Stanley-2008-PRL,Zhou-2008-PRE}, which makes it difficult to determine a proper scaling range. According to Fig.~\ref{Fig:MFDCCA:Stock}(d), it is evident that the MF-X-DMA algorithms with $\theta=0$ and 1 significantly outperform the MF-X-DFA algorithm.

\section{Conclusion and discussion}
\label{S1:conclusion}

In this work, we have developed a class of MF-DCCA algorithms based on the detrending moving average analysis. The performances of the MF-X-DMA algorithms are compared with the MF-X-DFA method by extensive numerical experiments on pairs of time series generated from bivariate fractional Brownian motions, two-component autoregressive fractionally integrated moving average processes and binomial measures, which have theoretical expressions of the multifractal nature. In all cases, the scaling exponents $h_{xy}$ extracted from the MF-X-DMA and MF-X-DFA algorithms are very close to the theoretical values.

For bivariate fractional Brownian motions, the scaling exponent $h_{xy}$ of the cross-correlation is found to be independent of the cross-correlation coefficient $\rho$ between two time series. The MF-X-DFA and centered MF-X-DMA algorithms outperform the forward and backward MF-X-DMA algorithms. When $H_{xx} \neq H_{yy}$, the MF-X-DFA and centered MF-X-DMA algorithms show comparable performance. When $H_{xx} = H_{yy}$, the centered MF-X-DMA algorithm performs slight better than the MF-X-DFA algorithm. Our numerical experiments verified the validity of the bFBM generating algorithm \cite{Chan-Wood-1999-SC,Coeurjolly-Amblard-Achard-2010-EUSIPCO,Amblard-Coeurjolly-Lavancier-Philippe-2011-BSMF}. For two-component autoregressive fractionally integrated moving average processes, we also found that the MF-X-DFA and centered MF-X-DMA algorithms have comparative performance, while the forward and backward MF-X-DMA algorithms perform slightly worse.  All the four algorithms are able to correctly unveil the monofractal nature in the cross-correlations between the components of bFMBs and two-components ARFIMA processes. For binomial measures, the forward MF-X-DMA algorithm exhibits the best performance, the centered MF-X-DMA algorithms performs worst, and the backward MF-X-DMA algorithm outperforms the MF-X-DFA algorithm when the moment order $q<0$ and underperforms when $q>0$.

In all the three mathematical models, the relation $h_{xy}=(h_{xx}+h_{yy})/2$ has been confirmed for all the four algorithms, where $h_{xy}$, $h_{xx}$ and $h_{yy}$ are estimated scaling exponents. Previous works have shown that the MF-DFA and MF-DMA algorithms are able to give nice estimates for univariate signals, that is, $h_{xx}\approx H_{xx}$ and $h_{yy}\approx H_{yy}$. It follows immediately that $h_{xy}\approx(H_{xx}+H_{yy})/2$. Combining the theoretical fact that $H_{xy}\approx(H_{xx}+H_{yy})/2$, we obtain $H_{xy}\approx h_{xy}$.
For monofractal time series, extensive numerical experiments unveiled that the performance of the DMA algorithms is comparable to the DFA algorithm and the centered DMA algorithm performs slightly better than DFA under certain situations \cite{Grech-Mazur-2005-APPB,Xu-Ivanov-Hu-Chen-Carbone-Stanley-2005-PRE,Bashan-Bartsch-Kantelhardt-Havlin-2008-PA}. This explains our numerical results on bFBMs and two-component ARFIMA processes. For multifractal measures generated from the $p$-model, the backward MF-DMA algorithm performs best \cite{Gu-Zhou-2010-PRE}, which explains our findings for MF-DCCA algorithms.

We applied these algorithms to the returns and volatilities of two US stock market indexes. For the returns, the centered MF-X-DMA algorithm gives the best estimates of $h_{xy}(q)$ since its $h_{xy}(2)$ is closest to 0.5, and the MF-X-DFA algorithm has the second best performance. For the volatilities, the forward and backward MF-X-DMA algorithms give similar results, while the centered MF-X-DMA and the MF-X-DFA algorithms fails to extract rational multifractal nature. These two applications are interesting since they showed that the choice of algorithms are automatic although we do not know which one should be used before analysis. The key message of our work is that we should use all the four algorithms and compare the results to make a choice.

\begin{acknowledgments}
We are grateful to Jean-Fran\c{c}ois Coeurjolly for providing the $R$ code to generate bivariate fractional Brownian motions and to Gao-Feng Gu and Qun-Zhi Zhang for invaluable discussions. This work was partially supported by the National Natural Science Foundation of China under grant no. 11075054 and the Fundamental Research Funds for the Central Universities.
\end{acknowledgments}

\appendix

\section{Higher-dimensional MF-X-DMA}

In this work, we have focused on time series analysis. It is easy to generalize the one-dimensional MF-X-DMA algorithms to higher dimensions. The higher-dimensional MF-X-DMA algorithms are closely related to the MF-X-DFA algorithms \cite{Zhou-2008-PRE}, the DMA algorithms \cite{Carbone-2007-PRE}, and the MF-DMA algorithms \cite{Gu-Zhou-2010-PRE} in higher dimensions.

Consider two physical quantities in $d$-dimension: $\{X(i_1,\cdots,i_d)\}$ and $\{Y(i_1,\cdots,i_d)\}$, where $i_j=1,2,\cdots,N_j$ for $j=1,2,\cdots,d$. Before proceeding, we need to construct the difference matrixes $\{x(i_1,\cdots,i_d)\}$ and $\{y(i_1,\cdots,i_d)\}$ of $X$ and $Y$. For simplicity, we denote $Z\in\{X,Y\}$ and $z$ as the corresponding difference matrix, which are related by the following equation:
\begin{equation}
  Z(i_1,\cdots,i_d) = \sum_{j_1=1}^{i_1}\cdots\sum_{j_d=1}^{i_d} z(i_1,\cdots,i_d).
  \label{Eq:Z:sum:z}
\end{equation}
The matrix $z$ is expressed as a square block matrix of size $2^d$, whose block is $z(i_1,\cdots,i_d)$, where the intervals $I_j=i_j$ or $[1:i_j-1]$ for $j=1,2,\cdots,d$. In real-world applications, we can focus on $d=2$ and $d=3$.

For the two-dimensional case $d=2$, the four blocks are $z(i_1,i_2)$, $z(i_1-1,i_2)$, $z(i_1,1:i_2-1)$, and $z(1:i_1-1,1:i_2-1)$.
According to Eq.~(\ref{Eq:Z:sum:z}), we have
\begin{eqnarray}
  Z(i_1,i_2) &=& z(i_1,i_2)+\sum_{j=1}^{i_1-1} z(j,i_2) \nonumber \\
             & & +\sum_{j=1}^{i_2-1} z(i_2,j)+\sum_{j_1=1}^{i_1-1}\sum_{j_2=1}^{i_2-1} z(j_1,j_2).
  \label{Eq:Z:sum:z:2}
\end{eqnarray}
Since
\begin{eqnarray}
  &&\sum_{j_1=1}^{i_1-1}\sum_{j_2=1}^{i_2-1} z(j_1,j_2) = Z(i_1-1,i_2-1),  \nonumber \\
  &&\sum_{j=1}^{i_1-1} z(j,i_2) = Z(i_1-1,i_2)-Z(i_1-1,i_2-1),  \nonumber \\
  &&\sum_{j=1}^{i_2-1} z(i_2,j) = Z(i_1,i_2-1)-Z(i_1-1,i_2-1),  \nonumber
\end{eqnarray}
it follows that
\begin{eqnarray}
  z(i_1,i_2) &=& Z(i_1,i_2)+Z(i_1-1,i_2-1) \nonumber \\
             & & -Z(i_1-1,i_2)-Z(i_1,i_2-1),
  \label{Eq:z:diff:Z:2}
\end{eqnarray}
where $Z(i,j)\triangleq 0$ if $i\times j=0$.

For the three-dimensional case $d=3$, the eight blocks are $z(i_1,i_2,i_3)$, $z(1:i_1-1,i_2,i_3)$, $z(i_1,1:i_2-1,i_3)$, $z(i_1,i_2,1:i_3-1)$, $z(1:i_1-1,1:i_2-1,i_3)$, $z(1:i_1-1,i_2,1:i_3-1)$, $z(i_1, 1:i_2-1,1:i_3-1)$, and $z(1:i_1-1,1:i_2-1,1:i_3-1)$. We can derive similarly as the two-dimensional case that
\begin{eqnarray}
  z(i_1,i_2,i_3) &=& Z(i_1,i_2,i_3) -Z(i_1-1,i_2-1,i_3-1) \nonumber \\
                 & & +Z(i_1-1,i_2-1,i_3)+Z(i_1-1,i_2,i_3-1) \nonumber \\
                 & & +Z(i_1,i_2-1,i_3-1)-Z(i_1-1,i_2,i_3) \nonumber \\
                 & & -Z(i_1,i_2-1,i_3)-Z(i_1,i_2,i_3-1),
  \label{Eq:z:diff:Z:3}
\end{eqnarray}
where $Z(i,j,k)\triangleq 0$ if $i \times j \times k =0$. When $i_3=1$, Eq.~(\ref{Eq:z:diff:Z:3}) reduces to Eq.~(\ref{Eq:z:diff:Z:2}).

The algorithm of $d$-dimensional multifractal detrending cross-correlation analysis is described as follows.

{\em{Step 1}}. For each quantity $z=x$ or $z=y$, determine the moving averages $\widetilde{Z}(i_1,\cdots,i_d)$, where $s_j\leqslant{i_j} \leqslant {N_j-\lfloor{(s_j-1)\theta_1}\rfloor}$  and $\{\theta_j\}$ are the position parameters with the values varying in the range $[0,1]$ for $j=1,2,\cdots,d$. For each point located at $(i_1,\cdots,i_d)$ in the $d$-dimensional space, we extract a sub-matrix $z(k_1,\cdots,k_d)$ with size $s_1\times\cdots\times{s_d}$ from the matrix $z$, where
$k_j\in[i_j-\lceil{(s_j-1)(1-\theta_j)}\rceil,i_j+\lfloor{(s_j-1)\theta_j}\rfloor]\triangleq[m_{j,1},m_{j,2}]$. We calculate the cumulative sums $Z'(k_1,\cdots,k_d)$ of the points within the box:
\begin{equation}
  Z'(k_1,\cdots,k_d)=\sum_{\ell_1=m_{1,1}}^{k_1}\cdots\sum_{\ell_1=m_{d,1}}^{k_2} {z(\ell_1,\cdots,\ell_d)},
\end{equation}
and the moving average $\widetilde{Z}(i_1,\cdots,i_d)$ at location $(i_1,\cdots,i_d)$ is calculated as follows,
\begin{equation}
  \widetilde{Z} =\frac{1}{s_1{\cdots}s_d} \sum_{k_1=m_{1,1}}^{m_{1,2}}\cdots\sum_{k_1=m_{d,1}}^{m_{d,2}} Z'(k_1,\cdots,k_d).
\end{equation}

{\em{Step 2}}. For each quantity, calculate the cumulative sums $Q(i_1,\cdots,i_d)$ in a sliding window with size $s_1\times\cdots\times{s_d}$, where $s_j\leqslant{i_j}\leqslant{N_j-\lfloor{(s_j-1)\theta_j}\rfloor}$. For each point located at $(i_1,\cdots,i_d)$, we have
\begin{equation}
  Q = \sum_{k_1=i_1-s_1+1}^{i_1}\cdots\sum_{k_d=i_d-s_d+1}^{i_d} z(k_1,\cdots,k_d).
  \label{Eq:2D:MFXDMA:Q}
\end{equation}

{\em{Step 3}}. Detrend the matrix by removing the moving average function $\widetilde{Z}(i_1,\cdots,i_d)$ from $Q(i_1,\cdots,i_d)$, and obtain the residual matrix $\epsilon_z(i_1,i_2)$ as follows,
\begin{equation}
  \epsilon^z(i_1,\cdots,i_d)=Q(i_1,\cdots,i_d)-\widetilde{Z}(i_1,\cdots,i_d),
  \label{Eq:2D:MFXDMA:Epsilon}
\end{equation}
where $s_j\leqslant{i_j}\leqslant{N_j-\lfloor{(s_j-1)\theta_j}\rfloor}$.

{\em{Step 4}}. Each residual matrix $\epsilon_z(i_1,\cdots,i_d)$ is partitioned into $N_{s_1}\times\cdots\times{N_{s_d}}$ disjoint boxes of the same size $s_1\times\cdots\times{s_d}$, where $N_{s_j}=\lfloor{(N_j-s_j(1+\theta_j))/s_j}\rfloor$. Each box can be denoted by
$\epsilon^z_{v_1,\cdots,v_d}$ for $v_j=1,\cdots,N_{s_j}$ such that $\epsilon^z_{v_1,\cdots,v_d}(k_1,\cdots,k_d)=\epsilon^z(l_{v_1}+k_1,\cdots,l_{v_d}+k_d)$ for $1\leqslant{k_j}\leqslant{s_j}$,
where $l_{v_j}=v_js_j$. The cross-correlation between $X$ and $Y$ in each box is calculated as follows:
\begin{eqnarray}
  F_{v_1,\cdots,v_d}&=&\frac{1}{s_1\dots{s_d}}\sum_{k_1=1}^{s_1}\cdots\sum_{k_d=1}^{s_d}\epsilon^x_{v_1,\cdots,v_d}(k_1,\cdots,k_d) \nonumber\\
  &&\times\epsilon^y_{v_1,\cdots,v_d}(k_1,\cdots,k_d).
  \label{Eq:2D:MFXDMA:Fv}
\end{eqnarray}

{\em{Step 5}}. The $q$th order overall detrending cross-correlation function $F_q(n)$ is calculated as follows:
\begin{equation}
  [F_q(s)]^q = \frac{1}{N_{s_1}{\cdots}N_{s_d}}\sum_{v_1=1}^{N_{s_1}}\cdots\sum_{v_d=1}^{N_{s_d}} |F_{v_1,\cdots,v_d}|^{q/2},
  \label{Eq:2D:MFXDMA:Fq}
\end{equation}
where $s^2=\sum_{j=1}^d s_j^2/d$ and $q$ can take any real values except for $q=0$. When $q=0$, we have
\begin{equation}
  \ln[F_0(n)] = \frac{1}{N_{s_1}{\cdots}N_{s_d}}\sum_{v_1=1}^{N_{s_1}}\cdots\sum_{v_d=1}^{N_{s_d}} \ln|F_{v_1,\cdots,v_d}|,
  \label{Eq:2D:MFXDMA:Fq0}
\end{equation}
according to L'H\^{o}spital's rule.

{\em{Step 6}}. Varying the box sizes $s_j$, we are able to determine the power-law relation between the fluctuation function ${F_q(s)}$ and the scale $s$,
\begin{equation}
  F_q(s)\sim{n}^{h(q)}.
  \label{Eq:2D:MFXDMA:hq}
\end{equation}

In real-world applications, one usually uses $s_1=\cdots=s_d=s$. When $N_{s_j}\neq (N_j-s_j(1+\theta_j))/s_j$, one needs to start from different directions as for the DFA algorithm \cite{Kantelhardt-KoscielnyBunde-Rego-Havlin-Bunde-2001-PA} or uses a random algorithm \cite{Ji-Zhou-Liu-Gong-Wang-Yu-2009-PA}. In addition, the box-by-box procedure is crucial for multifractal analysis, which was shown for high-dimensional MF-DFA \cite{Gu-Zhou-2006-PRE} and MF-DMA \cite{Gu-Zhou-2010-PRE}. However, the ``traditional'' procedure works well for high-dimensional fractals \cite{Carbone-2007-PRE}.

\begin{widetext}

\section{MATLAB codes for MF-X-DMA}

\lstset{breaklines}
\lstset{language=MATLAB}
\begin{lstlisting}

%% main function
function [Fxxq, Fxyq, Fyyq, s] = ZQJIANG_MFXDMA_1D(x, y, theta, q, s)
if nargin < 5
    L = length(x);
    s = [];i = 1.3;
    while round(10^i) <= L/4
        s = [s round(10^i)];
        i = i + 0.1;
    end
    clear L i
end
if nargin < 4, q = 2;end % DMA
if nargin < 3, theta = 0.5;end
% x y must be transversal vector
[i, j] = size(x);if i > 1 && j == 1, x = x';end
[i, j] = size(y);if i > 1 && j == 1, y = y';end
clear i j
Fxxq = zeros(length(q), length(s));
Fxyq = Fxxq;
Fyyq = Fxxq;
for i = 1:length(s)
    x_re = myfun_MA(x, s(i), theta);
    y_re = myfun_MA(y, s(i), theta);
    [Fxxq(:,i), Fxyq(:,i), Fyyq(:,i)] = myfun_Fq(x_re, y_re, s(i), q);
end
end
%% estimating the moving average residuals
function ts_re = myfun_MA(ts, s, theta)
% estimate moveing average
ts = cumsum(ts);
N = length(ts);
A = zeros(s, N-s+1);
for k = 1:s
    A(k, :) = ts(k:N-s+k);
end
MA = mean(A);
clear A
% moving average
ts_re = ts(1+floor((s-1)*theta):length(ts)-ceil((s-1)*(1-theta))) - MA;
% If the residuals can not be completely covered by the series, we can cover the series from both sides.
N = length(ts_re);
n = fix(N/s);
ls = N-n*s;
if ls ~= 0
    ts_re1 = ts_re(1:n*s); ts_re2 = ts_re(ls+1:N);
    ts_re = [ts_re1 ts_re2];
end
end
%% estimating the fluctuation scaling function
function [Fxxq, Fxyq, Fyyq] = myfun_Fq(x_re, y_re, s, q)
n = length(x_re)/s;
X = reshape(x_re, s, n);
Y = reshape(y_re, s, n);
Fxx = mean(abs(X).*abs(X));
Fxy = mean(abs(X).*abs(Y));
Fyy = mean(abs(Y).*abs(Y));
Fxxq = zeros(length(q), 1);
Fxyq = zeros(length(q), 1);
Fyyq = zeros(length(q), 1);
for k = 1:length(q)
    if q(k) ~= 0
        Fxxq(k,1) = (mean(Fxx.^(q(k)/2))).^(1/q(k));
        Fxyq(k,1) = (mean(Fxy.^(q(k)/2))).^(1/q(k));
        Fyyq(k,1) = (mean(Fyy.^(q(k)/2))).^(1/q(k));
    elseif q(k) == 0
        Fxxq(k,1) = exp(0.5*mean(log(Fxx)));
        Fxyq(k,1) = exp(0.5*mean(log(Fxy)));
        Fyyq(k,1) = exp(0.5*mean(log(Fyy)));
    end
end
end
\end{lstlisting}

\newpage

\section{MATLAB codes for MF-X-DFA}

\lstset{breaklines}
\lstset{language=MATLAB}
\begin{lstlisting}
%% main function
function [Fxxq, Fxyq, Fyyq, s] = ZQJIANG_MFXDFA_1D(x, y, q, order, s)
if nargin < 5
    L = length(x);
    s = [];i = 1.3;
    while round(10^i) <= L/4
        s = [s round(10^i)];
        i = i + 0.1;
    end
    clear L i
end
if nargin < 4, order = 1;end
if nargin < 3, q = 2;end % DFA
x = cumsum(x);
y = cumsum(y);
N = length(x);
Fxxq = zeros(length(q),length(s));
Fxyq = zeros(length(q),length(s));
Fyyq = zeros(length(q),length(s));
for i = 1:length(s)
    n = fix(N/s(i));
    ls = N-n*s(i);
    if ls ~= 0
        x1 = x(1:n*s(i));x2 = x(ls+1:N); rsx = [x1 x2];
        y1 = y(1:n*s(i));y2 = y(ls+1:N); rsy = [y1 y2];
        n = 2*n;
    else
        rsx = x;
        rsy = y;
    end
    X = reshape(rsx,s(i),n);
    Y = reshape(rsy,s(i),n);
    X = X';
    Y = Y';
    Fxx = zeros(1, n);
    Fxy = Fxx;
    Fyy = Fxx;
    for j = 1:n
        [Fxx(j), Fxy(j), Fyy(j)] = myfun_LocalResiduals(1:s(i), X(j,:), Y(j,:), order);
    end
    for k = 1:length(q)
        if q(k) ~= 0
            Fxxq(k,i) = (mean(Fxx.^(q(k)/2))).^(1/q(k));
            Fxyq(k,i) = (mean(Fxy.^(q(k)/2))).^(1/q(k));
            Fyyq(k,i) = (mean(Fyy.^(q(k)/2))).^(1/q(k));
        elseif q(k) == 0
            Fxxq(k,i) = exp(0.5*mean(log(Fxx)));
            Fxyq(k,i) = exp(0.5*mean(log(Fxy)));
            Fyyq(k,i) = exp(0.5*mean(log(Fyy)));
        end
    end
end
end
%% Detrend the local trends
function [Rxx, Rxy, Ryy] = myfun_LocalResiduals(t, x, y, n)
% n = 1 X = [x.^n;x.^(n-1)];
% n = 2 X = [x.^n;x.^(n-1);x.^(n-2)];
% n = 3 X = [x.^n;x.^(n-1);x.^(n-2);x.^(n-3)];
% n = 4 X = [x.^n;x.^(n-1);x.^(n-2);x.^(n-3);x.^(n-4)];
if n == 1
    Z = [t.^n;t.^(n-1)];
elseif n == 2
    Z = [t.^n;t.^(n-1);t.^(n-2)];
elseif n == 3
    Z = [t.^n;t.^(n-1);t.^(n-2);t.^(n-3)];
elseif n == 4
    Z = [t.^n;t.^(n-1);t.^(n-2);t.^(n-3);t.^(n-4)];
end
A = x/Z;
B = y/Z;
Rxx = mean(abs(x-A*Z).*abs(x-A*Z));
Rxy = mean(abs(x-A*Z).*abs(y-B*Z));
Ryy = mean(abs(y-B*Z).*abs(y-B*Z));
end
\end{lstlisting}
\end{widetext}
\bibliography{E:/Papers/Auxiliary/Bibliography}

\end{document}